\documentclass{aa}
\usepackage[varg]{txfonts}

\usepackage{graphicx}   
\usepackage{adjustbox}
\usepackage{natbib}
\bibpunct{(}{)}{;}{a}{}{,} 
\usepackage{soul}
\usepackage{xcolor}
\usepackage{hyperref}
\hypersetup{colorlinks=true, citecolor=blue, linkcolor=blue}

\newcommand{\orcid}[1]{\protect\href{https://orcid.org/#1}{\protect\includegraphics[width=10pt]{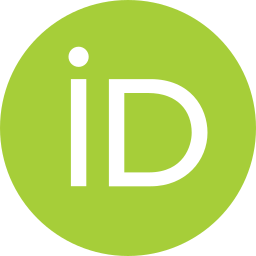}}}

\newcommand{\msun}{\mbox{$M_{\odot}$}}
\newcommand{\msunyr}{\mbox{$M_{\odot}\,{\rm yr^{-1}}$}}
\newcommand{\kms}{\mbox{${\rm km\,s^{-1}}$}}
\newcommand{\mvir}{\mbox{$M_{\rm vir}$}}
\newcommand{\mstar}{\mbox{$M_{\rm star}$}}
\newcommand{\mhalo}{\mbox{$M_{\rm halo}$}}
\newcommand{\cmq}{\mbox{$\,{\rm cm^{-3}}$}}
\newcommand{\nH}{\mbox{$n_{\rm H}$}}
\newcommand{\eff}{\mbox{$\varepsilon_{\rm ff}$}}
\newcommand{\tff}{\mbox{$t_{\rm ff}$}}
\newcommand{\racc}{\mbox{$r_{\rm acc}$}}
\newcommand{\rsonic}{\mbox{$r_{\rm sonic}$}}
\newcommand{\dx}{\mbox{$\Delta x$}}
\newcommand{\dxsq}{\mbox{$\Delta x^2$}}
\newcommand{\dxmin}{\mbox{$\Delta x_{\rm min}$}}
\newcommand{\dxminsq}{\mbox{$\Delta x^2_{\rm min}$}}

\definecolor{pink}{RGB}{255,51,153}

\begin{document} 

   \title{Impact of star formation models on the growth of simulated galaxies at high redshifts}

   \author{Cheonsu Kang\inst{1}\thanks{E-mail:astro.ckang@gmail.com}\orcid{0009-0003-1180-4979}
          \and
          Taysun Kimm\inst{1}\thanks{E-mail:tkimm@yonsei.ac.kr}\orcid{0000-0002-3950-3997}
          \and
          Daniel Han\inst{1}\orcid{0000-0002-2624-3129}
          \and
          Harley Katz\inst{2}\orcid{0000-0003-1561-3814}
          \and
          Julien Devriendt\inst{3}\orcid{0000-0002-8140-0422}
          \and
          Adrianne Slyz\inst{3}
          \and
          Romain Teyssier\inst{4}\orcid{0000-0001-7689-0933}
          }

   \institute{Department of Astronomy, Yonsei University, 50 Yonsei-ro, Seodaemun-gu, Seoul 03722, Republic of Korea
   \and
   Department of Astronomy and Astrophysics, University of Chicago, Chicago, Illinois 60637, USA
   \and
   Sub-department of Astrophysics, University of Oxford, Denys Wilkinson Building, Keble Road, Oxford OX1 3RH, UK
   \and
   Department of Astrophysical Sciences, Princeton University, Peyton Hall, Princeton NJ 08544, USA}
            
   \date{Received XXX; accepted YYY}

   \abstract
   {Star formation is a key process that governs the baryon cycle within galaxies, however, the question of how it controls their growth remains elusive due to modeling uncertainties. To understand the impact of star formation models on galaxy evolution, we performed cosmological zoom-in radiation-hydrodynamic simulations of a dwarf dark matter halo, with a virial mass of $\mvir \sim 10^9\,\msun$ at $z=6$. We compared two different star formation models: a multi-freefall model combined with a local gravo-thermo-turbulent condition and a more self-consistent model based on a sink particle algorithm, where gas accretion and star formation are directly controlled by the gas kinematics. As the first study in this series, we used cosmological zoom-in simulations with different spatial resolutions and found that star formation is more bursty in the runs with the sink algorithm, generating stronger outflows than in the runs with the gravo-thermo-turbulent model. The main reason for the increased burstiness is that the gas accretion rates on the sinks are high enough to form stars on very short timescales, leading to more clustered star formation. As a result, the star-forming clumps are disrupted more quickly in the sink run due to more coherent radiation and supernova feedback. The difference in burstiness between the two star formation models becomes even more pronounced when the supernova explosion energy is artificially increased. Our results suggest that improving the modeling of star formation on small, sub-molecular cloud scales can significantly impact the global properties of simulated galaxies.}

   
   \titlerunning{Bursty star formation with sinks}

   \maketitle



\section{Introduction}
Star formation is the fundamental process that determines the properties of galaxies. It affects the distribution of gas by removing cold gas from star-forming dense clumps and ejecting matter into the interstellar medium (ISM) and intergalactic medium through various forms of stellar feedback. The gas ejected from the galaxy then recollapses and forms stars as it cools, balancing heating and cooling and regulating the growth of galaxies \citep[for a review, see][]{Somerville2015}.

Star formation is known to be a slow process \citep[e.g.,][]{Krumholz2007slow, McKee2007}. The Schechter form of galaxy mass functions, which differs from the power law predicted for dark matter halos \citep[DMHs,][]{Press1974}, already suggested that star formation is inefficient in both low-mass and high-mass galaxies \citep{Cole2001, Bell2003, Read2005}. Indeed, by comparing the ratio of stellar mass to DMH mass using the abundance-matching technique \citep{Moster2010}, it has been shown that the conversion efficiency of gas to stars is generally low on galactic scales \citep{Behroozi2013}. Studies based on the rotation curves of dwarf galaxies \citep{Read2017} have also reached the same conclusion, suggesting that only a few percent of the universal baryons are converted into stars in halos with virial masses $\mvir\lesssim10^{10.5}\,\msun$. In low-mass systems where the gravitational potential is shallow, the explosion of supernovae (SNe) is thought to suppress star formation \citep{Dekel1986}; whereas in massive galaxies, the accretion energy of supermassive black holes may have driven galactic winds and offset the cooling flow \citep{Binney1995}.

However, it has been a challenge for numerical simulations to reproduce the observed properties of dwarf and/or $L_\star$ galaxies \citep{Naab2017}. This has proven to be difficult especially in the center of galaxies, where gas flows in from all directions, forming too many stars and making the galaxy very compact \citep[e.g.,][]{Joung2009}. One culprit is excessive cooling during SN explosions, as the cooling length tends to be underresolved in galactic-scale simulations \citep{Katz1992, Kimcg2015}. Several methods have been proposed to overcome this by explicitly incorporating the momentum-conserving phase \citep{Kimm2014, Hopkins2014, Oku2022}. For example, \citet{Kimm2015} have found that simulations with their mechanical SN feedback suppress star formation in halos with mass $\mvir\lesssim10^{10}\,\msun$ better than those with thermal or kinetic SN feedback. Applying the same scheme to the \textsc{arepo} code, \citet{Smith2018}  showed that the Kennicutt-Schmidt relation \citep{Schmidt1959, Kennicutt1998} is indeed better reproduced than with classical thermal or kinetic schemes \citep[see also][]{Rosdahl2017}. However, the authors also point out that in dense regions of galaxies, SNe alone cannot regulate star formation as observed \citep{Rosdahl2017, Smith2019}. Instead, \citet{Keller2014} argued that hot superbubbles driven by clustered young stars can produce stronger outflows if thermal conduction is properly taken into account. Radiation from massive stars is also thought to be an important source of pressure that can disrupt giant molecular clouds (GMCs) in star-forming galaxies by driving the expansion of HII bubbles \citep{Matzner2002, Krumholz2007magneto, Kimjg2018}. Indeed, cosmological simulations have found that star formation can be further suppressed if the radiation feedback from massive stars is included in addition to SN explosions \citep{Wise2012, Rosdahl2015}. Finally, cosmic rays accelerated by SNe have recently emerged as a potential solution, as they can produce continuous galactic winds without artificially increasing the SN energy or randomly distributing the SN explosion \citep{Jubelgas2008, Simpson2016, Farcy2022}.

Since these feedback energies come from massive stars, bursty star formation histories (SFHs) can, in principle, help explain the observational properties of galaxies. For example, \citet{Faucher2018} argued that massive outflows can be better explained when star formation is modeled in a bursty way. It can also flatten the central density profile in dwarf galaxies by non-adiabatically changing the orbits of dark matter particles \citep{Governato2010, Teyssier2013, Chan2015}; however, this does not seem to happen in a galaxy with a smooth SFH \citep{Vogelsberger2014}. The detection of UV-bright galaxies at $z\gtrsim 10$ in the recent JWST survey has challenged the standard cosmological model \citep[e.g.,][]{Finkelstein2022, Naidu2022, Harikane2024}, but using FIRE-2 cosmological zoom-in simulations, \citet{Sun2023} argued that the abundance of galaxies with $M_{\rm UV}<-20$ is naturally explained by rapidly time-varying SFHs. The same conclusion has also been reached with an empirical approach, where high UV variability or large scatter in the SFR can increase the abundance of UV-bright galaxies \citep{Mirocha2023, Shen2023}, highlighting the importance of accurately modeling the burstiness of star formation.

In numerical simulations of galaxy formation, the star formation process is often implemented as subgrid physics, assuming a Schmidt law \citep{Schmidt1959}:
\begin{equation}
    \frac{d M_\star}{dt} = \eff\frac{M_{\rm gas}}{\tff},
    \label{eq:schmidt}
\end{equation}
where $M_\star$ is the stellar mass, $M_{\rm gas}$ is the gas mass, and \tff\ is the free-fall time. The most uncertain parameter is the SFE per free-fall time (\eff), which is typically a few percent at densities relevant to GMC scales \citep{Krumholz2007slow, Evans2009, Utomo2018}. \citet{KM05} showed analytically that \eff\ varies with the virial parameter and the turbulent Mach number of GMCs. \citet{HC11} and \citet{PN11} further considered the multi-freefall collapse and magnetic fields for star formation, respectively. Comparing high-resolution magnetohydrodynamic simulations with observations, \citet{FK12} argued that interstellar turbulence is the primary source controlling the star formation rate (SFR). Indeed, \citet{Braun2015} simulated an isolated disk galaxy in a $10^{12}\,\msun$ DMH with both single and multi-freefall collapse models,  finding that the observed relationship between SFR surface density and gas surface density is better reproduced with multi-freefall models. This would indicate that the gas inside star-forming clouds collapses on different timescales.

Identifying potential star formation sites presents its own set of challenges. Stars are known to form in regions of dense gas, and as such, a density threshold is commonly used as a prerequisite for star formation. This threshold can be used in conjunction with temperature or convergent gas flow criteria \citep{Cen1992}. In addition to these criteria, \citet{Cen1992} also introduced dynamical restrictions, requiring the gas to have a cooling time shorter than the dynamical time and to be Jeans-unstable. Building upon the previous work, and motivated by the fact that stars form in GMCs, some studies have assumed that star formation takes place in molecular hydrogen-rich gas and have shown that the model reproduces the molecular Schmidt-Kennicutt relation \citep{Robertson2008} or the observed transition from atomic to molecular hydrogen \citep{Gnedin2009}. To determine the most reliable way to model star formation, \citet{Hopkins2013} simulated two disk galaxies using seven different criteria, including those mentioned above. These authors showed that the simulated SFRs converge reasonably well when stellar feedback is included, with star formation occurring in self-gravitating regions emerging as the most physically motivated criterion. In a similar vein, \citet{Nunez-Castineyra2021} investigated the impact of multi-freefall star formation on the evolution of a Milky Way-like galaxy and showed that the detailed structure of the ISM depends on the interplay between star formation and feedback models. \citet{Girma2024} further argued that pressure support from magnetic fields can make the Schmidt-Kennicutt relation shallower, highlighting the importance of magnetic field dynamics in regulating star formation rates within galaxies.

To mitigate the uncertainties associated with \eff\ and star formation criteria, it thus seems natural to attempt to model the formation and growth of star particles directly. In simulations where the ISM structure is resolved in some detail, this can be done by seeding density peaks within gas clumps and allowing these seeds to grow by subsequent accretion of inflowing gas. We note that this approach provides a more physically motivated and self-consistent alternative to multi-freefall models, improving our understanding of formation and evolution of galaxies.
Because of its advantages, the sink particle algorithm is widely used to build a framework for high-resolution simulations, ranging from GMC scales (e.g., \citealt{Vazquez2007}; STARFORGE, \citealt{Grudic2021}) to galactic scales (SILCC, \citealt{Gatto2017}; TIGRESS, \citealt{Kim2017}), and to explore the global impact of star formation in controlled experiments. For example, using such a model, \citet{Tress2020} simulated an M51-like galaxy at sub-parsec resolution and showed that the overall star formation rate is mostly controlled by the self-regulated properties of the ISM rather than by the large-scale gas flows triggered by the interaction with a companion galaxy. Based on the TIGRESS framework, \citet{Moon2022} simulated nuclear rings with inflowing gas using sink particles with a uniform spatial resolution of 4 pc. They found that the depletion time, which is inversely proportional to the SFR, agrees well with the predictions of self-regulation theory \citep{Ostriker2010}. More recently, the STARFORGE model was  applied to cosmological simulations to resolve the formation and evolution of individual stars on sub-pc scales \citep{Hopkins2024}. In this work, we further study the impact of the sink particle algorithm on SFHs by simulating a dwarf galaxy at high redshift in cosmological settings, and compare its results to a subgrid model based on gravo-thermodynamic gas properties.

This paper is organized as follows. In Section 2, we describe the initial conditions, numerical methods, input physics, and the two star formation models used in our simulations. The main results of the radiative hydrodynamics simulations of a dwarf galaxy, which focus on comparing the burstiness and feedback strength in the two star formation models, are presented in Section 3. In Section 4, we discuss the intrinsic model difference, the effect of Type II SN explosion energy, and spatial resolution on our conclusions. Finally, we summarize our results in Section 5. Tests of the sink particle algorithm against singular isothermal sphere collapse and turbulent GMC simulations are presented in the appendix.

\section{Simulations}
To study the relationship between burstiness of SFH and strength of SN feedback, we performed both idealized and cosmological simulations using an adaptive mesh refinement (AMR) code, \textsc{ramses-rt} \citep{Teyssier2002, Rosdahl2015}. The Euler equations were solved using the Harten-Lax-van Leer Contact (HLLC) scheme \citep{Toro1994}. The Poisson equation was solved using a multigrid method \citep{Guillet2011}. The radiative transfer calculation was performed using a fluid of photon approach with the M1 closure \citep{Levermore1984} and the global Lax-Friedrich scheme. A reduced speed of light approximation was employed where $c$, the speed of light, was set to a hundredth of its true value. A Courant factor of 0.7 was adopted. Photoionization, direct momentum transfer from photon absorption, and non-thermal pressure due to trapped infrared (IR) photons were modeled on the basis of eight photon groups ([0.1, 1.0), [1.0, 5.6), [5.6, 11.2), [11.2, 13.6), [13.6, 15.2), [15.2, 24.59), [24.59, 54.42), [54.42, $\infty$) in units of eV), as described in Table 2 of \citet{Kimm2017}. These values were coupled with the non-equilibrium chemistry of seven primordial species ($\rm H_2$, HI, HII, HeI, HeII, HeIII, $\rm e^-$), allowing for more accurate calculations of cooling rates \citep{Rosdahl2013, Katz2017}. A redshift-dependent uniform UV background radiation from \citet{HM12} was included in the calculation of the photoionization and heating with a self-shielding factor $\exp(-\nH/0.01\,\cmq)$, following \citet{Rosdahl2013}. At $T>10^4\,{\rm K}$, cooling from metal species was considered by interpolating predefined tables as a function of gas density and temperature \citep[\texttt{cc07} model,][]{Ferland1998}. For lower temperatures, metal cooling due to fine structure transitions was modeled according to \citet{Rosen1995}. For dust, we assumed a constant dust-to-metal ratio of 0.4 at temperature below $10^5\, \rm K$.

\subsection{Stellar feedback}
In our cosmological simulations, we considered several forms of stellar feedback: photo-ionization heating and direct radiation pressure from massive stars, non-thermal IR pressure, Type II SN explosions, and mass-dependent explosions of Pop III stars. Interested readers are referred to \citet{Kimm2017} for details, but we describe the main features of the models here for completeness.

We injected the photon flux from each star particle as a function of its age, metallicity and mass. For Pop III stars, we took the lifetimes and photon production rates from \citet[][Tables 4--5]{Schaerer2002}. For Pop II stars, we employed the Binary Population and Spectral Synthesis model \citep[BPASS, v2.0,][]{Stanway2016} with a maximum stellar mass of 100\,\msun\ and a Kroupa initial mass function \citep{Kroupa2002}. Among the eight photon groups, Lyman continuum photons ($\lambda<912$ \r{A}) ionize neutral hydrogen, heating the gas and forming HII bubbles around the stars. The absorption of these photons by the neutral ISM, or the absorption of non-ionizing radiation by dust can also transfer momentum to their surroundings. In addition, we include the non-thermal pressure from trapped IR photons, although its effect is negligible due to the low dust optical depth in our early low-mass galaxies.

Type II SN feedback is implemented using a mechanical feedback scheme \citep{Kimm2014,Kimm2015}. Assuming that stars more massive than $8\, \msun$ explode as core-collapse SNe, a simple stellar population of mass $100\, \msun$ would host one SN explosion. This means that each star particle of mass $500\, \msun$ explodes five times. The lifetime of the Type II SN progenitor is sampled between $\approx 4$ and 50 Myr \citep[see Fig. 2 of][]{Kimm2015} and, thus, the exact location and epoch of the SN explosion is different for each SN progenitor. Each explosion imparts radial momentum to 48 neighboring cells (if AMR levels are uniform), and the final amount of radial momentum depends on the gas properties of the SN site \citep[e.g.,][]{Thornton1998,Kimcg2015}, characterized as
\begin{equation}
    P_{\rm SN}=2.5\times10^5\, \kms\, E^{16/17}_{51} n^{-2/17}_{\rm H} Z'^{-0.14},
    \label{eq:Psn}
\end{equation}
where $E_{51}$ is the Type II SN energy in units of $10^{51}\, \rm erg$, $n_{\rm H}$ is the hydrogen number density in units of \cmq, and $Z'=\rm max[0.01,\ Z/Z_\odot]$ is the gas phase metallicity, normalized to the solar value ($Z_\odot=0.02$). The fiducial value of $E_{51}$ is 1, and we run additional simulations with $E_{51}=5$ to study how more energetic explosions affect the evolution of galaxies in the simulations with two different star formation models. The chemical yield of the SN ejecta is assumed to be $\eta_{\rm Z}=0.075$.

For Pop III stars, we release different energies and metal yields based on the progenitor mass. This is done by categorizing Pop III stars into four groups: normal Type II SN \citep[$11-20\ \msun$,][]{Nomoto2006}, hypernova \citep[$20-40\ \msun$,][]{Nomoto2006}, or pair instability SN \citep[$140-260\ \msun$,][]{Heger2002}. We assume that neither energy nor metals are released from Pop III stars if their mass does not fall in any of these three mass ranges.

\subsection{Star formation models}
In this section, we describe the differences between the two different star formation models, a subgrid scheme based on local gravo-thermo-turbulent (GTT) conditions \citep{Kimm2017} and a more direct model based on a sink particle algorithm \citep{Bleuler2014}. To minimize the potential differences due to random sampling of Pop III stellar masses, we apply the same Pop III formation scheme in cosmological simulations as in Kimm et al. (2017). The formation criteria for Pop III are essentially identical to those of the GTT model, but the stellar metallicity is assumed to be lower than $Z_{\rm star} \le 2\times10^{-8}$ \citep[e.g.,][]{Omukai2005}.

\subsubsection{A subgrid model based on gravo-thermo-turbulent conditions}
In cosmological simulations with finite resolution, star formation has typically been modeled using a simple density criterion and a fixed \eff\ \citep[e.g.,][]{Vogelsberger2013,Dubois2014,Kimm2014}. The GTT model is designed to incorporate the multi-freefall nature of the star formation process, with particular applications to high-resolution cosmological simulations such as NewHorizon \citep{Dubois2021} and SPHINX \citep{Rosdahl2018}.

\begin{figure}
    \includegraphics[width=\linewidth]{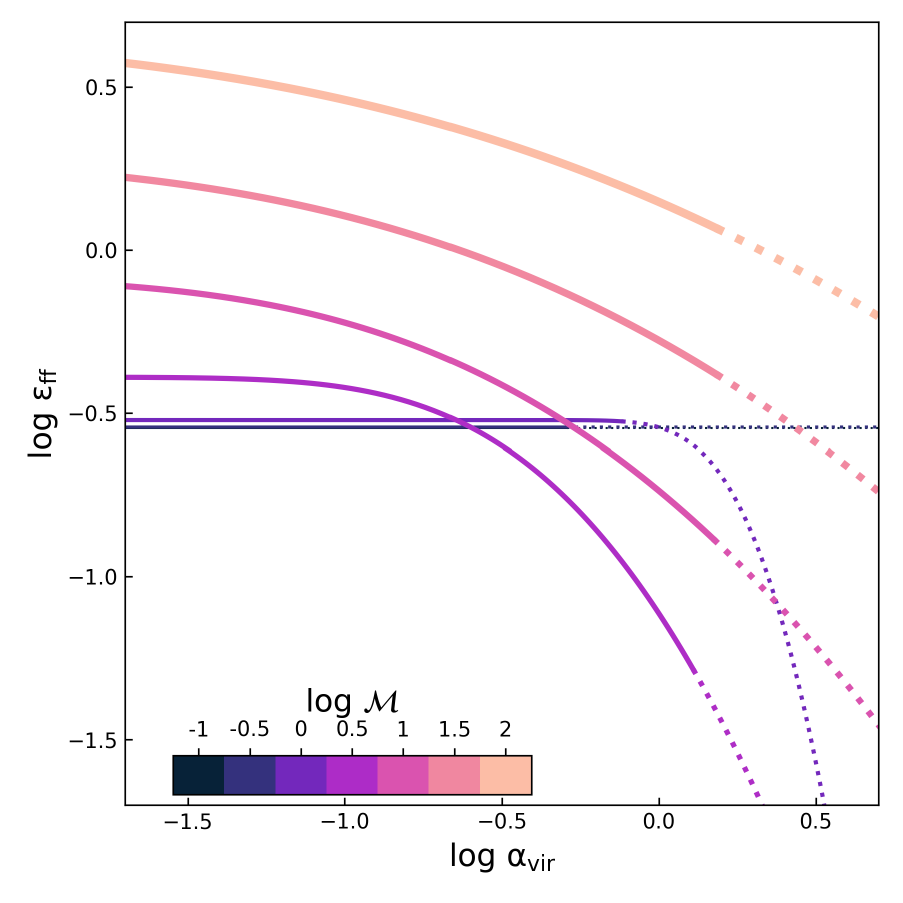}
    \caption{Star formation efficiency per free-fall time ($\varepsilon_{\rm ff}$) as a function of virial parameter ($\alpha_{\rm vir}$) used in the local gravo-thermo-turbulent (\texttt{GTT}) subgrid model. We plot $\varepsilon_{\rm ff}$ for seven different Mach numbers ($\mathcal{M}$) equally spaced in logarithmic intervals from 0.1 to 100. The curves for the two smallest $\mathcal{M}$ overlap so only six curves are shown on the plot. The dotted lines indicate the region where star formation is inhibited according to Eq.~\eqref{eq:alpha_trunc}.}
    \label{fig:eff_map}
\end{figure}

The GTT model assumes that the central cell in which a star particle is to form, together with its six neighbor cells, forms a spherical cloud. Based on a Schmidt law, the amount of newly formed stars is computed as in Eq.~\eqref{eq:schmidt}. The main difference from previous simple models used in galaxy formation simulations is the use of the variable \eff, which is calculated using the multi-freefall star formation theory \citep{PN11, FK12}, as
\begin{equation}
    \eff=\frac{\varepsilon_{\rm ecc}}{2\phi_{\rm t}} \rm exp \left(\frac{3}{8}\sigma^2_{\rm s} \right) \left[2-\rm erfc \left(\frac{\sigma^2_{\rm s}-s_{\rm crit}}{\sqrt{2\sigma^2_{\rm s}}}\right)\right].
    \label{eq:eff}
\end{equation}
Here $\varepsilon_{\rm ecc}$ is the maximum fraction of gas that can accrete without being blown away by protostellar feedback, which is set to 0.5.
$1/\phi_{\rm t} \approx 0.57$ accounts for the uncertainty in the free-fall time measurement \citep{KM05}, and is chosen as the best-fit value from \citet{FK12}. The remaining parameters are
\begin{equation}
    \sigma^2_{\rm s}=\ln \left(1+b^2\mathcal{M}^2\right), \, s_{\rm crit}=\ln \left(0.067\theta^{-2}\alpha_{\rm vir} \mathcal{M}^2\right),
\end{equation}
where $\sigma_{\rm s}$ is the standard deviation of the logarithmic density contrast, assuming that the probability distribution function (PDF) of the density of a star-forming cloud can be well described by a log-normal distribution, and $s_{\rm crit}$ is the critical (minimum) density required for the gas to collapse. The turbulence parameter $b$ accounts for the mixing ratio of different turbulence modes. We use a value of 0.4 for $b$, which corresponds to an approximate 3:1 mixture of solenoidal (divergence-free) and compressive (curl-free) modes \citep[e.g.,][]{Federrath2010turb,Orkisz2017}. $\theta=0.33$ is a proportionality constant between the diameter of a spherical cloud and the post-shock thickness \citep{PN11}. \eff\ then simply becomes a function of two parameters representing the local gas properties, the turbulent Mach number $\mathcal{M}$ ($\equiv \sigma_{\rm gas}/c_{\rm s}$) and the local virial parameter $\alpha_{\rm vir}$, as
\begin{equation}
\alpha_{\rm vir}=\frac{5(\sigma^2_{\rm gas}+c^2_s)}
    {\pi G \rho \dxsq}=\frac{5(\mathcal{M}^2+1)}{\pi^2}\left(\frac{\lambda_{\rm J}}{\dx}\right)^2,
\end{equation}
where $\sigma_{\rm gas}$ is the gas turbulent velocity, $c_{\rm s}$, $\rho$ and $\lambda_{\rm J}$ are the sound speed, the total (gas plus star and dark matter) density, and the thermal Jeans length of the central cell, respectively. The unresolved turbulent velocity at the scale of a central cell (\dx) is computed as the gas velocity dispersion from the central and six adjacent cells. This is done after subtracting the mean, symmetric divergence, and rotational velocities from the velocity of each cell to extract the random motion. The relationship between \eff\ and $\alpha_{\rm vir}$ at seven different $\mathcal{M}$ is shown in Fig.~\ref{fig:eff_map}. We note that if a cloud is highly turbulent and strongly bound, \eff\ can be greater than 1. However, we find that \eff\ in the \texttt{GTT} runs ranges from 0.05 to 0.37, with a clear peak around 0.3, regardless of the spatial resolution.

For a cloud to collapse and form stars, the gravitational force must be stronger than the thermal plus turbulent pressure gradient. Based on this argument, star formation is triggered when the local cell size is larger than the turbulent Jeans length \citep{Bonazzola1986}, as
\begin{equation}
    \lambda_{\rm J,turb}=\frac{\pi \sigma^2_{\rm gas}+\sqrt{\pi^2 \sigma^4_{\rm gas} + 36\pi G \rho c^2_s \dxsq}}{6G \rho \dx} \le \dx.
    \label{eq:lambda_jt}
\end{equation}
This condition can also be expressed with parameters representing local gas properties, as
\begin{equation}
    \alpha_{\rm vir} \le \frac{15}{\pi^2} \frac{\mathcal{M}^2+1}{\mathcal{M}^2+3}, \, {\rm or} \ \frac{\lambda_{\rm J}}{\dx} \le \sqrt{\frac{3}{\mathcal{M}^2+3}}.
    \label{eq:alpha_trunc}
\end{equation}
\st{\bf Note that} For strong (weak) turbulence, this effectively reduces the range over which we apply the GTT subgrid model to $\alpha_{\rm vir}<1.52$ ($\alpha_{\rm vir}<0.5$, respectively), as indicated by the dotted lines in Fig.~\ref{fig:eff_map}.
By construction, coarse cells cannot satisfy Eqs.~\eqref{eq:lambda_jt} and \eqref{eq:alpha_trunc}, because refinement is triggered when the thermal Jeans length is less than $8\dx$ and therefore star formation can only occur in maximally refined cells.

We also require a convergent gas flow condition toward the central cell. In addition, we examine the Eq.~\eqref{eq:lambda_jt} criterion only for cells with densities greater than $\nH\ge 100\,\cmq$ to reduce the computational cost of searching for neighboring cells of potential star-forming sites. In practice, stars form at much higher densities, and Eq.~\eqref{eq:lambda_jt} imposes a stricter criterion than others, thereby controlling the overall star formation process.

The minimum mass of the star particles is chosen so that each particle hosts five SNe events, which means $m_{*,\rm min}=500\, \msun$. In principle, it is possible to increase the stellar mass resolution even further, but we defer this improvement to a later investigation, because in this case the initial mass function must be properly sampled to account for the continuum emission from individual stars \citep{Kroupa2013}, as done in \citet{Andersson2023} for example. We allow for the formation of more massive star particles, with masses integer multiples of $m_{*,\rm min}$, by sampling a Poisson distribution \citep{Rasera2006}. The probability of creating a more massive star increases with increasing gas mass and \eff. For example, in the cosmological zoom-in simulations presented in this work (Sect.~\ref{sec:30}), some Pop II stars in our simulations with the lowest spatial resolution are as massive as $10\,m_{*,\rm min}$, but $\sim 85\%$ of the total stellar mass is composed of minimum-mass star particles and this fraction increases with resolution.

\subsubsection{Star formation model based on sink particles}\label{sec:222}
Sink particles were first introduced by \citet{Bate1995} to study accretion onto protostellar systems in a Lagrangian code. The idea was to replace crowded gas particles in high-density regions, which cause time steps to shorten, with a single non-gaseous particle representing a protostar. The low-mass seed star then accretes surrounding gas, in contrast to most subgrid models for star formation where the star particle mass is fixed at formation.

The formation criteria for such sink particles are not significantly different from other subgrid star formation models, as both involve forming within dense clumps. A widely used sink formation criterion is the Truelove condition \citep[e.g.,][]{Krumholz2004, Federrath2010sink, Bleuler2014}. \citet{Truelove1997} suggested that the thermal Jeans length should be resolved with at least four cells in gravitationally collapsing regions, otherwise artificial fragmentation may occur. 
This condition gives the density threshold 
\begin{equation}
    \rho_{\rm Tr}=\frac{\pi c^2_s}{16G \Delta x_{\rm min}^2},
\end{equation} 
where \dxmin\ is the size of the finest cell. If $\dxmin=1\, \rm pc$, the value of $\rho_{\rm Tr}$ for the typical ISM gas with a temperature of 30 K and a mean molecular weight of 2.273 is about $5.7\times10^{-22}\, {\rm g \,cm^{-3}}$, or $\nH \approx 260\,\cmq$. In contrast, \citet{Gong2013} used the density profile of a self-gravitating isothermal sphere \citep{Larson1969, Penston1969} $\rho_{\rm LP}(r)=\frac{8.86 c^2_s}{4 \pi G r^2}$ and set the threshold at $r=0.5\dxmin$. The corresponding density threshold is
\begin{equation}
    \rho_{\rm LP}(0.5\dxmin)=\frac{8.86 c^2_s}{\pi G \dxminsq},
\end{equation}
which is $\approx 14.4$ times higher than $\rho_{\rm Tr}$.

Once a sink is formed, its growth depends on how the accretion rate is calculated. There exists an analytic solution for spherical accretion onto a point mass, first described by \citet{Hoyle1939} and \citet{Bondi1952}, as\footnote{Strictly speaking there are two analytic solutions: one for the Bondi case with the accreting point mass at rest and the other for the Hoyle-Lyttleton case with the accreting mass moving highly supersonically with respect to the surrounding gas \citep[see][for more details]{Edgar2004}. Eq.~\eqref{eq:Bondi} is an educated interpolation between these two regimes which is not always accurate in the transonic regime \citep[see e.g.,][]{Beckmann2018}.} 
\begin{equation}
\dot M_{\rm acc}=4 \pi \rho_\infty r^2_{\rm BH} \sqrt{c^2_\infty+v^2_\infty},
\label{eq:Bondi}
\end{equation}
where $\rho_\infty,\, c_\infty,\, v_\infty$ are the gas density, the sound speed, and the relative velocity to the ambient medium far from the accreting mass, respectively. The Bondi-Hoyle radius $r_{\rm BH}=GM_\star/(c^2_\infty+v^2_\infty)$ is the maximal impact parameter gas can have in order to be accreted. Bondi-Hoyle accretion was used in the pioneer study of \citet{Bate1995} and later implemented in an adaptive mesh refinement (AMR) code by \citet{Krumholz2004}.
However, as pointed out by these authors, it is not trivial to measure $\rho_\infty,\, c_\infty$, and $v_\infty$ in the simulation, which motivates the search for alternative methods to model the accretion rates.
For example, \citet{Federrath2010sink} employed a scheme based on a density threshold that allows for mass growth when the gas density of cells within the accretion zone exceeds $\rho_{\rm Tr}$. The mass accreted onto a sink particle is calculated such that the remaining gas density inside the accretion zone equals $\rho_{\rm Tr}$. This approach ensures that the average density surrounding a sink particle remains similar to $\rho_{\rm Tr}$ over a long period of time. While this method is simple and effective in preventing artificial fragmentation, it does not take into account the motion of the gas, which can be as important as the gas density in determining the accretion rate. \citet{Gong2013} introduced a flux accretion scheme that takes into account the velocity of the infalling gas when computing the accretion rate. In this scheme, the accretion rate is set to the inflowing mass flux at the boundary of the accretion zone. \citet{Bleuler2014} modified this scheme by replacing the gas velocity with its relative velocity with respect to the sink particle to correct for the motion of this latter.
Using the divergence theorem, the accretion rate in this case can be written as
\begin{equation}
    \dot M_{\rm acc} = \oint \rho_{\rm gas} (\vec{v}_{\rm gas}-\vec{v}_{\rm sink}) \cdot \vec{dA} = \int \rho_{\rm gas} \vec{\bigtriangledown} \cdot (\vec{v}_{\rm gas}-\vec{v}_{\rm sink})\, dV,
\end{equation}
where $\rho_{\rm gas}$ and $\vec{v}_{\rm gas}$ are the density and velocity of the gas within the accretion zone, and $\vec{v}_{\rm sink}$ is the velocity of the sink particle. The first and second integrals represent the closed surface and volume integrals, respectively, over the accretion zone.
The three different accretion schemes are tested against the analytic solution at the center of a collapsing isothermal sphere, as given by \citet{Shu1977}, and all of them are found to reproduce the analytic solution well \citep{Krumholz2004, Federrath2010sink, Gong2013}.

In our simulations, a sink particle is formed at the density peak of a gas clump. We first identify gas clumps using isodensity contours \citep{Bleuler2015}, based on \textsc{clumpfind} from \citet{Williams1994}. A sink particle is then created if the following set of conditions are met:
\begin{itemize}
    \item[-] the peak of a clump is at the maximum AMR level and has a density higher than a given threshold,
    \item[-] a clump is virialized,
    \item[-] there is a net gas inflow ($\vec{\bigtriangledown} \cdot \vec{v} < 0$), and
    \item[-] there is no pre-existing sink particle located closer than one accretion radius.
\end{itemize}
For the density threshold, we adopt the \citet{Gong2013} criterion, $\rho_{\rm th}=\rho_{\rm LP}(0.5\dxmin)$. Since it is unlikely that all the clumps of gas in a galaxy have the same temperature, we calculate $\rho_{\rm th}$ on a cell-by-cell basis.

For the accretion radius, we take $\racc=2\dxmin$, as this provides a reasonable compromise between accuracy for the accretion rate on a singular isothermal sphere and the ability to capture local structures (see Sect.~\ref{sec:A1}). Each sink particle is surrounded by 257 cloud particles of equal mass. They are uniformly distributed throughout the accretion zone, positioned every 0.5\dxmin\ along the three axes. Cells containing one or more cloud particles are used to calculate the physical quantities of a parent sink particle, including accretion rate and velocity. This ensures that there are 54 maximally refined cells within the accretion zone. Each cell has a different weight depending on the number of cloud particles it contains. For comparison, the number of cloud particles for $\racc=\dxmin$ and $\racc=4\dxmin$ is 33 and 2109, respectively.

To calculate the accretion rate, we use both the flux-based and the Bondi-Hoyle accretion schemes. Following \citet{Bleuler2014}, we use the sonic radius as the switch between two schemes, such that the flux accretion scheme is activated when $\rsonic > \racc$ (see Sect.~\ref{sec:A1} for more details), where $\rsonic=GM_{\rm sink}/2c^2_{\rm s}$ is the sonic radius, and $M_{\rm sink}$ is the sink particle mass. \citet{Bleuler2014} introduced a correction factor that slightly modifies the flux accretion rate depending on the mean gas density within the accretion zone in order to keep the gas density similar to the threshold density for a long time and to avoid artificial fragmentation. However, we do not use it in this work because the corrected form slightly underestimates the accretion rate in the collapsing isothermal sphere test. For $\rsonic < \racc$, we use the Bondi-Hoyle accretion scheme. The accuracy of the two schemes is discussed further in Sect.~\ref{sec:A1}. The seed mass of the sink particle is set to $0.1\, \msun$, which is small enough that the sudden conversion of gas into a particle does not affect the local gas dynamics. Note that the hydrodynamic time step is also limited to prevent the sink mass from doubling in a single fine time step. To avoid having too many sink particles sharing an accretion zone, we allow sink particles to merge if their distances to one another becomes smaller than $2\,\racc$.

The formation of a star particle is triggered when a sink becomes more massive than a given threshold \citep[e.g.,][]{Han2022}. We set this mass threshold to $500\, \msun$, which is equal to $m_{*,\rm min}$ in the GTT model. A new star particle is created at the position of its parent sink particle, and the corresponding mass is subtracted from the sink. If there is a continuous supply of gas to the sink, multiple star particles can form from a single sink.

\section{Results}\label{sec:30}
In this section, we present results from cosmological zoom-in simulations performed using the two star formation models previously described. We assess whether bursty star formation can enhance feedback strength and regulate star formation. We quantify burstiness for each model and compare feedback strengths by analyzing when and where SN explosions occur and their collective effect on star-forming gas clumps and galactic outflows.

\begin{table}
\caption{Summary of the cosmological simulations at $z=6$.}
\label{table1}
\begin{adjustbox}{width=\columnwidth}
\centering
\begin{tabular}{l c c c c c}
\hline \hline
Simulation &  $\dxmin$  & $\dxmin$ & $E_{\rm SN}$ & $\mhalo$ &  $\mstar$\\ 
Label      &  [cpc]     &  [pc]     &  [$10^{51} \rm erg$] & [$10^9\,\msun$] & [$10^6\,\msun$] \\
\hline
\texttt{GTT-Fid}       & 4.8 & 0.7 & 1 & 1.22 & 2.08 \\
\texttt{GTT-LR2}       & 9.5 & 1.4 & 1 & 1.21 & 2.41 \\
\texttt{GTT-LR4}       & 19  & 2.8 & 1 & 1.18 & 2.62 \\
\texttt{GTT-LR8}       & 38  & 5.5 & 1 & 1.23 & 2.30 \\
\texttt{GTT-LR16}      & 76  & 11  & 1 & 1.24 & 5.35 \\
\texttt{GTT-LR8-E5}    & 38  & 5.5 & 5 & 1.24 & 1.21 \\
\texttt{SINK-Fid}      & 4.8 & 0.7 & 1 & 1.25 & 2.10 \\
\texttt{SINK-LR2}      & 9.5 & 1.4 & 1 & 1.25 & 3.48 \\
\texttt{SINK-LR4}      & 19  & 2.8 & 1 & 1.22 & 2.53 \\
\texttt{SINK-LR8}      & 38  & 5.5 & 1 & 1.25 & 2.59 \\
\texttt{SINK-LR16}     & 76  & 11  & 1 & 1.19 & 3.02 \\
\texttt{SINK-LR8-E5}   & 38  & 5.5 & 5 & 1.13 & 0.40 \\ 
\hline
\end{tabular}
\end{adjustbox}
\tablefoot{From left to right, each column indicates the simulation name, the size of the finest cell in comoving units, the size of the finest cell in physical units, the energy of individual Type II SN, the halo mass, and the stellar mass. All simulations have mass resolutions of $490\,\msun$ and $500\,\msun$ for dark matter and star particles, respectively.}
\end{table}

\begin{figure*}
    \includegraphics[width=\textwidth]{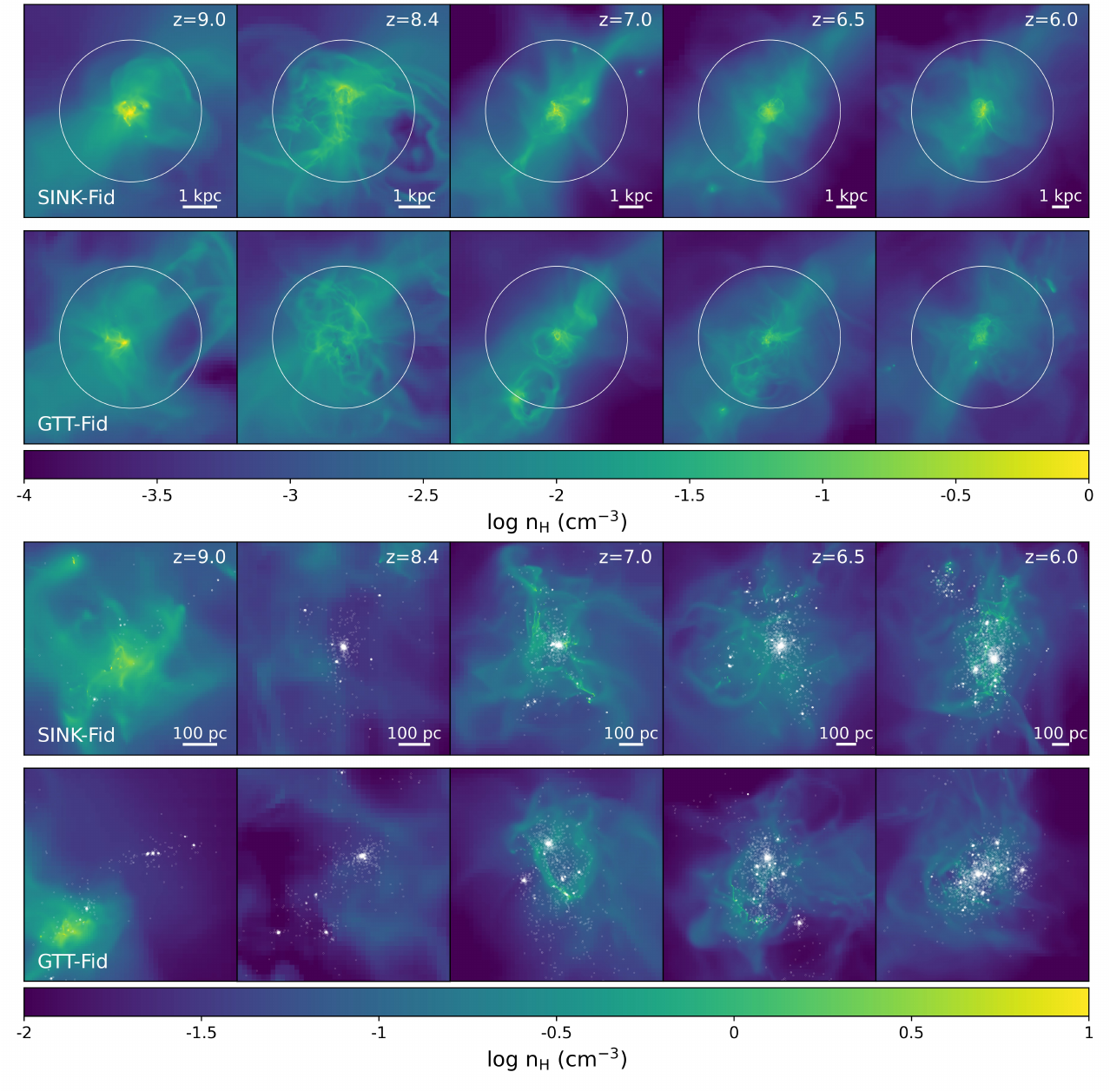}
    \caption{Projected images of the hydrogen number density around the main halo in our fiducial runs at five different epochs. The white circle in the upper panels represents the virial radius, and the length scale is shown as a white scale bar in the lower right corner of each panel. The lower panels show the blow-up images of the central halo, with a different color scale to the upper panels. Star particles are shown as white dots.}
    \label{fig:cosmo_map}
\end{figure*}

\subsection{Star formation history and burstiness}
\begin{figure}
    \includegraphics[width=\linewidth]{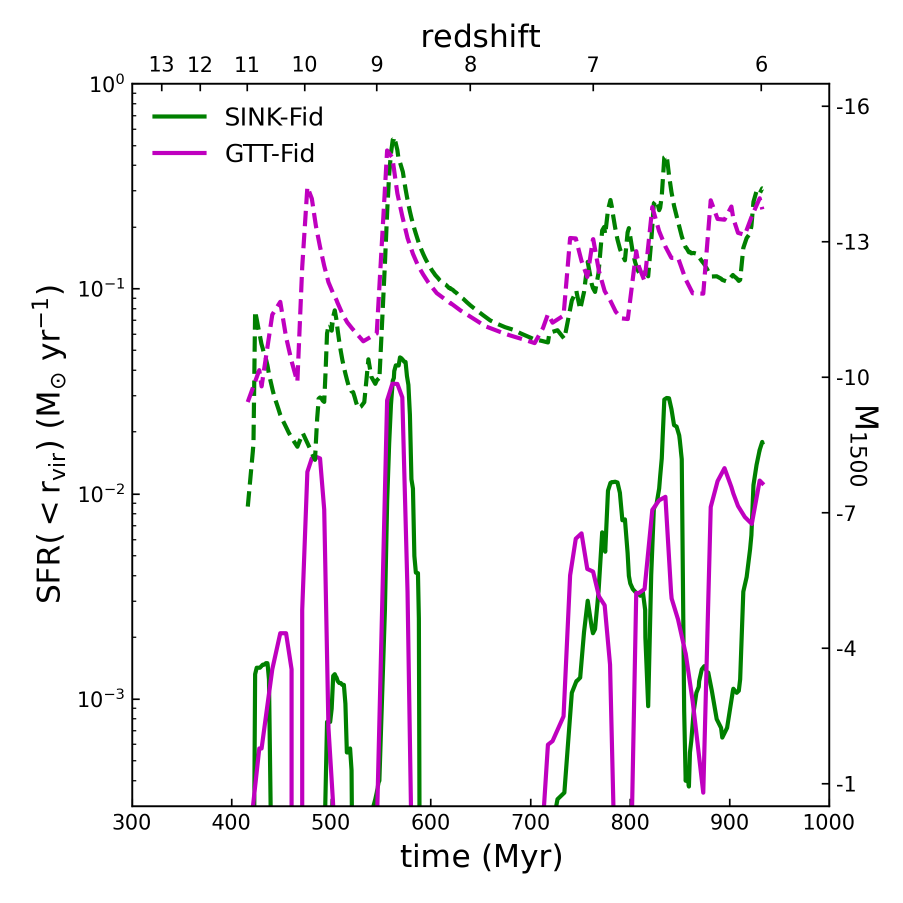}
    \caption{Star formation rates (solid lines) and intrinsic UV magnitudes measured at 1500 \r{A} ($M_{\rm 1500}$, dashed lines) measured within the main halo in our fiducial runs. Star formation rates are averaged over 20 Myr, and $M_{\rm 1500}$ are computed by integrating over the wavelength range between 1450 and 1550 \r{A}.}
    \label{fig:SFR}
\end{figure}

We begin by describing the merging history and SFH of our simulated halo of mass $10^9\,\msun$ at $z=6$ from the fiducial run with a resolution of 4.8 cpc (or 0.7 pc at $z=6$). In Fig.~\ref{fig:cosmo_map} we show the projected gas distributions of the most massive progenitor at five different epochs. At $z\approx9$ two galaxies embedded in DMHs of masses $10^8\, \msun$ and $3\times10^7\, \msun$ merge and undergo a starburst. The SFR rises up to $\approx 0.04\,\msunyr$ or $5\times10^{-8}\, \rm yr^{-1}$ for the specific SFR, as can be seen in Fig.~\ref{fig:SFR}. Subsequent SN explosions expel gas out of the galactic center and drive significant galactic outflows. As a result, star formation is quenched for $\sim 100\, {\rm Myr}$ until gas cools down and recollapses at $z\approx7.4$. Afterwards, star formation events become more episodic (peaks at $z\approx 7$, $6.5$ and $6$), as stellar feedback tends to disrupt the star-forming clouds. Although not explicitly shown, at $z=7$ two small halos cross the virial sphere and merge at $z\sim 6$ (with a total mass ratio of 6:1), but no dramatic starburst occurs. This is because the satellite galaxy is gas deficient at the time of the merger due to its own stellar feedback, as illustrated by the giant bubble near the virial sphere (upper middle panel for \texttt{GTT-Fid} in Fig.~\ref{fig:cosmo_map}). At $z=6$, the main simulated galaxy has a stellar mass of $2\times10^6\,\msun$ and a mean stellar metallicity of $0.005\,Z_\odot$. 

Figure~\ref{fig:SFR} compares star formation histories in the two runs: gravo-thermo-turbulent model (magenta) and sink particle algorithm (green). This figure also shows UV absolute magnitudes ($1450 < \lambda$/\r{A} $< 1550$; $M_{1500}$) for each galaxy obtained by interpolating SEDs from BPASS \citep{Stanway2016} to account for the age and metallicity, as well as the mass of each star particle. The AB magnitude system is used \citep{Oke1990} and dust absorption is ignored as the galaxy is very metal-poor. We find that the peaks of individual star formation events generally tend to be higher in \texttt{SINK-Fid} than in \texttt{GTT-Fid}. In the very early phase of its evolution ($z\gtrsim 10$), when the halo is less massive than $\lesssim 10^8\,\msun$ star formation is very stochastic and the reverse may happen, but as the halo becomes more massive, star formation histories become more bursty in the \texttt{SINK} model than in \texttt{GTT-Fid}. This can also be seen in $M_{1500}$ at lower redshifts, where the main galaxy peaks in the \texttt{SINK-Fid} run are twice as bright at 1500 \r{A} as those in the \texttt{GTT-Fid} run. As we will show later, the burstier nature of the galaxies simulated with the sink algorithm for star formation is also a characteristic of the \texttt{SINK} runs with different resolutions or feedback strengths.

To be more quantitative, we parameterize the burstiness of star formation by calculating the ratio, motivated by \citet{Caplar2019}:
\begin{equation}
    B=\frac{\sigma/\mu-1}{\sigma/\mu+1},
    \label{eq:B-param}
\end{equation}
where $\mu$ and $\sigma$ are the mean and standard deviation of the offset from the average star formation ($\Delta$). \citet{Caplar2019} assumed a power law of the form $d\mstar/dt \propto \mstar^\alpha$ to define the main sequence and measured $\Delta_{\rm MS}$  on a logarithmic scale ($\Delta_{\rm MS} \equiv \log({\rm SFR}/{\rm SFR_{MS}})$). However, this method cannot be applied to our dwarf galaxies because star formation is too stochastic to fit with a power law and is sometimes completely quenched.
Instead, we calculate the absolute offset ($\Delta =|{\rm SFR} - {\rm SFR_{avg}}|$) from an average SFR  of each run between $6<z<12$ (e.g.,  ${\rm SFR_{avg}}=0.005\,\msunyr$ for \texttt{SINK-Fid}) , and compute $\sigma$ and $\mu$ on a linear scale. Here the SFRs are averaged over 1 Myr as in \citet{Caplar2019}.
However, unlike \citet{Caplar2019} where $B$ is measured on the 30 Myr timescale, we measure it for $6<z<12$ and adopt the absolute difference    $|{\rm SFR} - {\rm SFR_{avg}}|$ to avoid $\mu$ being 0 by construction. This is slightly different from \citet{Caplar2019}, but we emphasize that the physical implication remains the same in the sense that large $\sigma/\mu$ refers to a more bursty SFH.
Since both $\mu$ and $\sigma$ are non-negative, $B$ ranges from -1 (constant signal, $\sigma/\mu=0$) to 1 (maximally bursty, $\sigma/\mu \rightarrow \infty$).
We find that $B=0.37$ and $B=0.21$ in the \texttt{SINK-Fid} and \texttt{GTT-Fid} runs respectively, confirming our inference from Fig.~\ref{fig:SFR} that star formation is indeed more bursty in \texttt{SINK-Fid}. These mean that the offset ($\Delta_{\rm MS}$) in the \texttt{SINK-Fid} run fluctuates more strongly ($\sim30\%$) than that in \texttt{GTT-Fid}, which we argue is significant.

\begin{figure}
    \includegraphics[width=\linewidth]{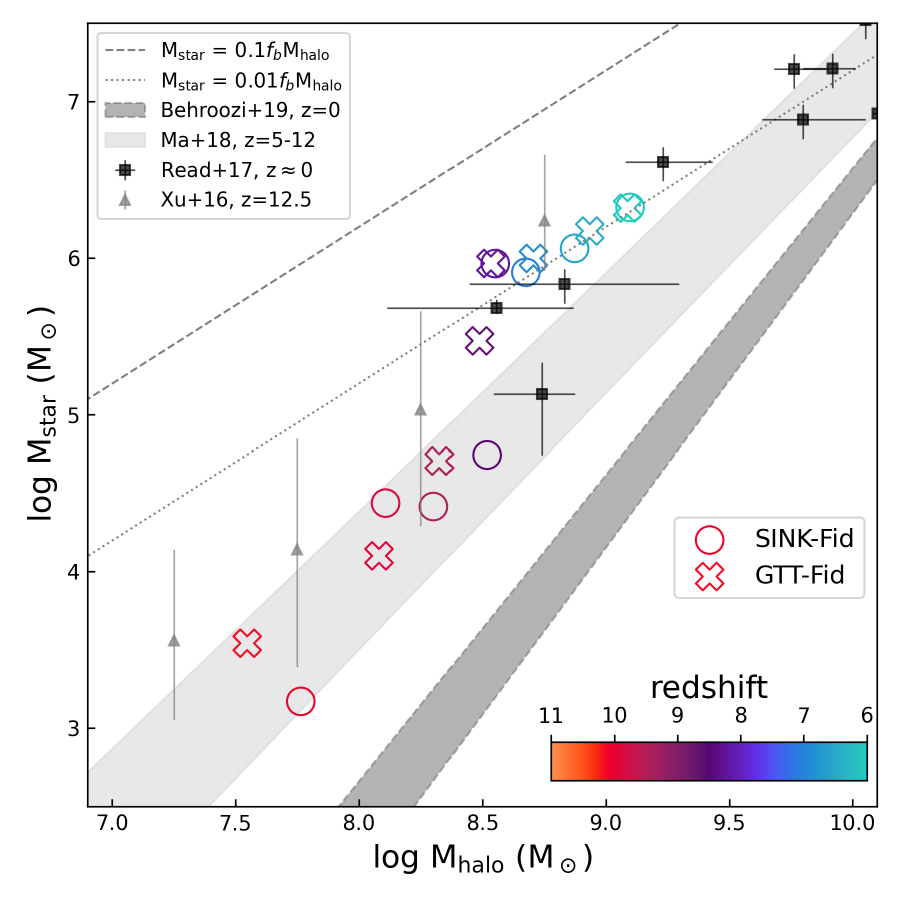}
    \caption{Galactic stellar mass as a function of DMH mass in our fiducial runs from $z=11$ to $6$. Different open symbols indicate the runs with different star formation models, with their colors representing the redshift. The dashed and dotted lines correspond to 10\% and 1\% gas-to-star conversion efficiencies, respectively. Also shown as dark and light shaded regions are the (extrapolated) relation at $z=0$ \citep[][]{Behroozi2019} and at $z=5$--$12$ \citep{Ma2018} with 68\% confidence intervals, respectively. The relation for the local field dwarfs is plotted as black squares \citep{Read2017}. The results of the Renaissance simulation are shown as gray triangles with vertical error bars indicating the 1$\sigma$ range \citep{Xu2016}. Note that the stellar mass referred to here is the current mass, inclusive of mass loss from SNe, not the total mass of stars originally formed.}
    \label{fig:SMHM}
\end{figure}

Figure~\ref{fig:SMHM} shows the stellar-mass-to-halo-mass relation for the main progenitor galaxies at different redshifts. Despite the fact that the burstiness of star formation is different, we find that the final stellar masses of the two runs at $z=6$ are nearly identical. For a given halo mass, the stellar masses in the \texttt{SINK-Fid} run are smaller most of the time, but during the merger-driven or local star formation events the \texttt{SINK-Fid} galaxy creates more stars than \texttt{GTT-Fid}. At $z>9$, both models convert 0.1\% of the total baryon into stars ($\mstar \approx 0.001 f_{\rm b}\, M_{\rm halo}$), and the fraction then increases up to one percent as the halo mass increases. A similar trend is found in the Renaissance simulation \citep{Xu2016}, although our predicted stellar masses are on average smaller by a factor of two. It is also interesting to note that our simulated galaxy masses are comparable to the stellar-mass-to-halo-mass estimates from local dwarf galaxies \citep{Read2017}. In contrast, the predicted stellar masses from our simulations are larger than those from the FIRE simulation \citep{Ma2018} or those inferred from the extrapolation of the local stellar mass-to-halo mass ratio obtained by the abundance matching technique \citep{Behroozi2019}.

The different amount of burstiness in the two star formation runs means that the mass-to-light ratio ($M_\star/L_{1500}$) of the galaxies may be different. Indeed, during active star formation phases, galaxies can become UV-bright in a manner not necessarily correlated with their overall stellar mass. To obtain the $M_\star/L_{1500}$ ratio, we calculate the total stellar mass within the virial sphere of the dark matter halo host and compare it with the total luminosity emitted at 1450 \r{A} $<\lambda <$ 1550 \r{A}.
We find that in \texttt{GTT-Fid}, $M_\star/L_{1500}$ varies from 0.06 to 2.9 $M_{\odot}/L_{\odot}$ at $6<z<12$, with an average of $0.93 \, M_{\odot}/L_{\odot}$, while this ratio turns out to be smaller ($0.71\,M_{\odot}/L_{\odot}$) in the \texttt{SINK-Fid} run.
We also measure that the minimum $M_\star/L_{1500}$ value in \texttt{SINK-Fid}, which occurs during the first starburst ($z\sim9$), is 1.5 times smaller than in \texttt{GTT-Fid}. This supports the claim of \citet{Sun2023} that a recent surge in star formation provides a plausible explanation for the presence of UV-bright galaxies detected by JWST at redshifts greater than 10 \citep[e.g.,][]{Finkelstein2022, Castellano2023,Harikane2024}, although a top-heavy initial mass function \citep{Yung2024} increased star formation efficiency during the earliest stages of galaxy formation \citep{Dekel2023, Qin2023}, accreting black holes \citep{Inayoshi2022} or non-standard cosmological models \citep{Hirano2024} can be considered as possible alternatives.

\subsection{Feedback strength}\label{sec:32}
\begin{figure}
    \includegraphics[width=\linewidth]{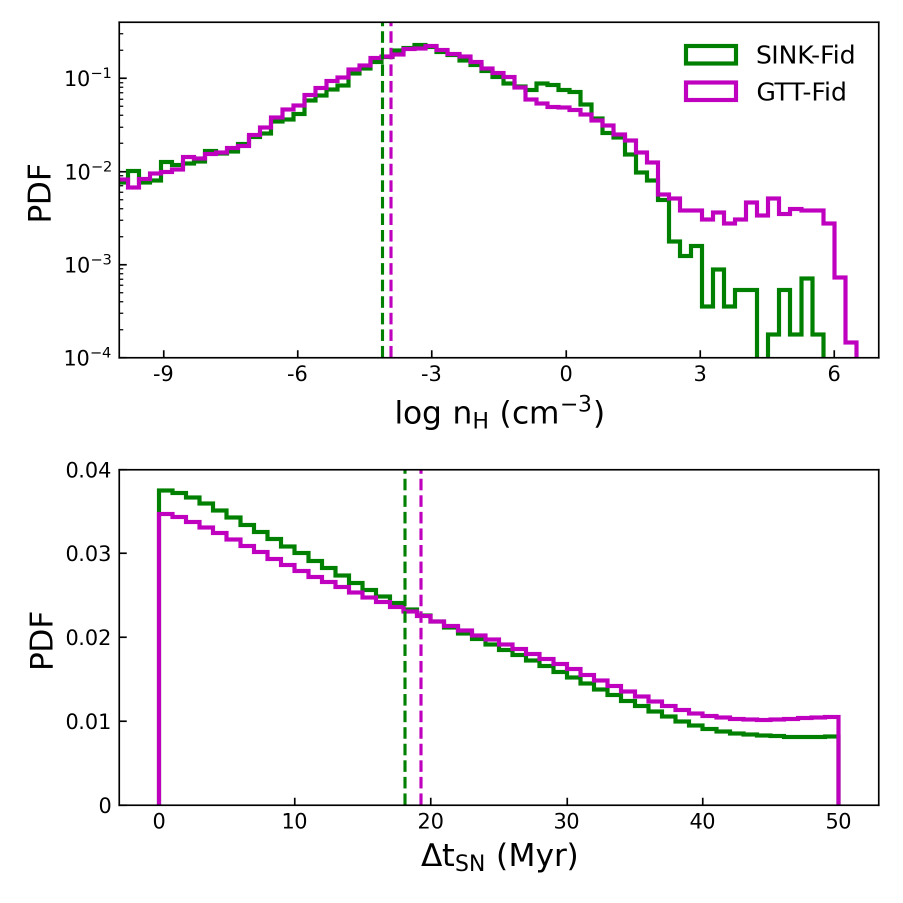}
    \caption{PDF of hydrogen number density (\nH) of SN host cells (top). The mean logarithmic density is shown as a vertical dashed line.  PDF of the time interval between two randomly selected SN explosions throughout the simulation (bottom). The vertical dashed line indicates the average time gap between two different SNe.}
    \label{fig:SN}
\end{figure}

To understand the impact of bursty star formation on the feedback strength, we present in Fig.~\ref{fig:SN} two different metrics associated with SN explosions. The top panel shows the hydrogen number density of the host cell in which the SNe explode. While gas metallicity distributions are almost indistinguishable between \texttt{GTT-Fid} and \texttt{SINK-Fid} (not shown), we see that the number of SNe exploding at the highest densities ($\log \nH \gtrsim 2.5$) is significantly reduced in the \texttt{SINK-Fid} model. As we will discuss later (Sect.~\ref{sec:51}), this is partly due to the fact that stars form in less dense environments in the \texttt{SINK} runs ($\left<\log \nH \right>=5.6$ vs $6.0$). Furthermore, we find that the average number of star particles younger than 3 Myr per cell (if present) is 3.8 and 2.5 in \text{SINK-Fid} and \texttt{GTT-Fid} respectively, indicating that stars form in a more clustered fashion and that radiation feedback is thus stronger in the \texttt{SINK} run. As a result, SNe explode at densities that are about a factor of two lower in \texttt{SINK-Fid} ($\log n_{\rm H} \sim -4.1$) than in \texttt{GTT-Fid}, and more momentum is injected into the ISM per SN explosion in \texttt{SINK-Fid} ($\left<P_{\rm SN}\right>\approx 1.8\times10^6\, \msun\ \kms$) than in \texttt{GTT-Fid} ($\left<P_{\rm SN}\right>\approx 1.3\times10^6\, \msun\ \kms$) in Eq.~\eqref{eq:Psn}. 

The lower panel of Fig.~\ref{fig:SN} shows how SN explosions are correlated in time. We computed the PDF of the time gap between two randomly selected SNe exploding in the zoom-in region. This simple calculation ignores the spatial distribution of the SNe, but we argue that time gap information is closely related to individual star formation events, since star formation in our dwarf galaxy is generally bursty and clustered. Figure~\ref{fig:SN} shows that SNe explode together rather than independently. This feature is more pronounced in \texttt{SINK-Fid}, where the stars form in a more clustered fashion. Although not dramatic, both density of SN host cell and time gap suggest that stellar feedback is stronger in \texttt{SINK-Fid}, compared to \texttt{GTT-Fid}.

\begin{figure}
    \includegraphics[width=\linewidth]{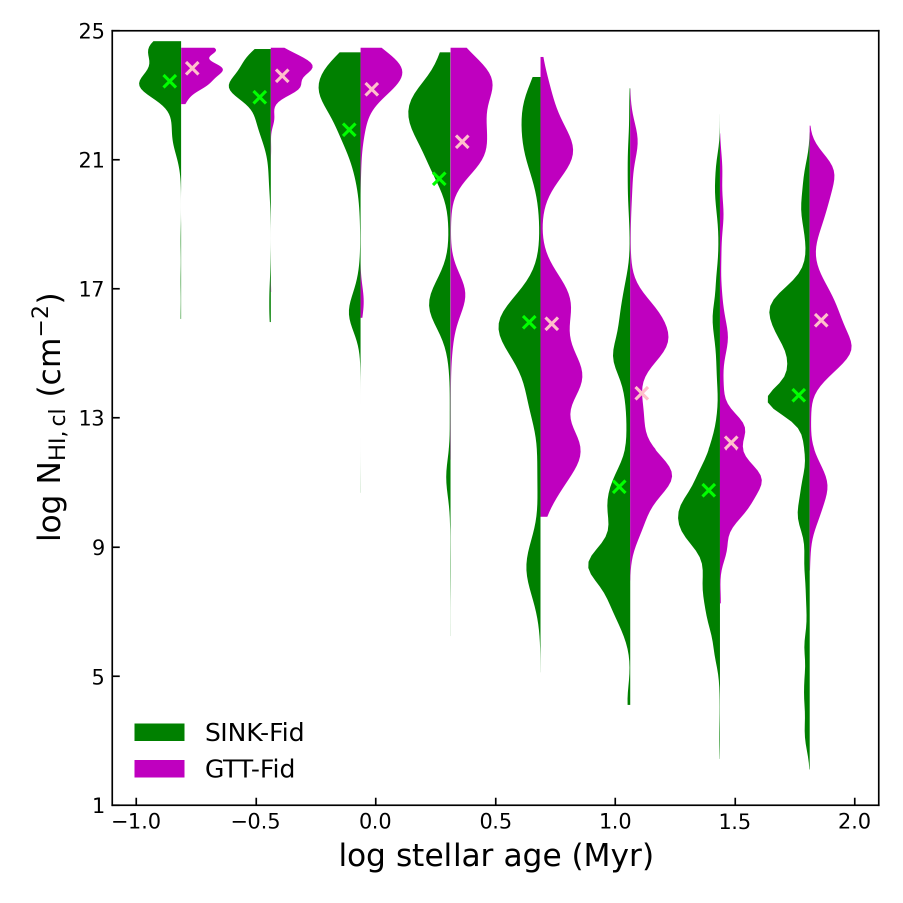}
    \caption{Distribution of the atomic hydrogen column density integrated over 50 pc from each star particle as a function of stellar age. We measure the column density around star particles within $0.2\, r_{\rm vir}$ in all snapshots. The column density is averaged over six directions, and the stellar age is uniformly binned on a logarithmic scale into 8 intervals from 0.1 to 100 Myr. The mean logarithmic column densities of each bin are indicated by crosses. Note that the column density decreases faster in the \texttt{SINK-Fid} run.} 
    \label{fig:violin}
\end{figure}

\begin{figure}[!ht]
    \includegraphics[width=\linewidth]{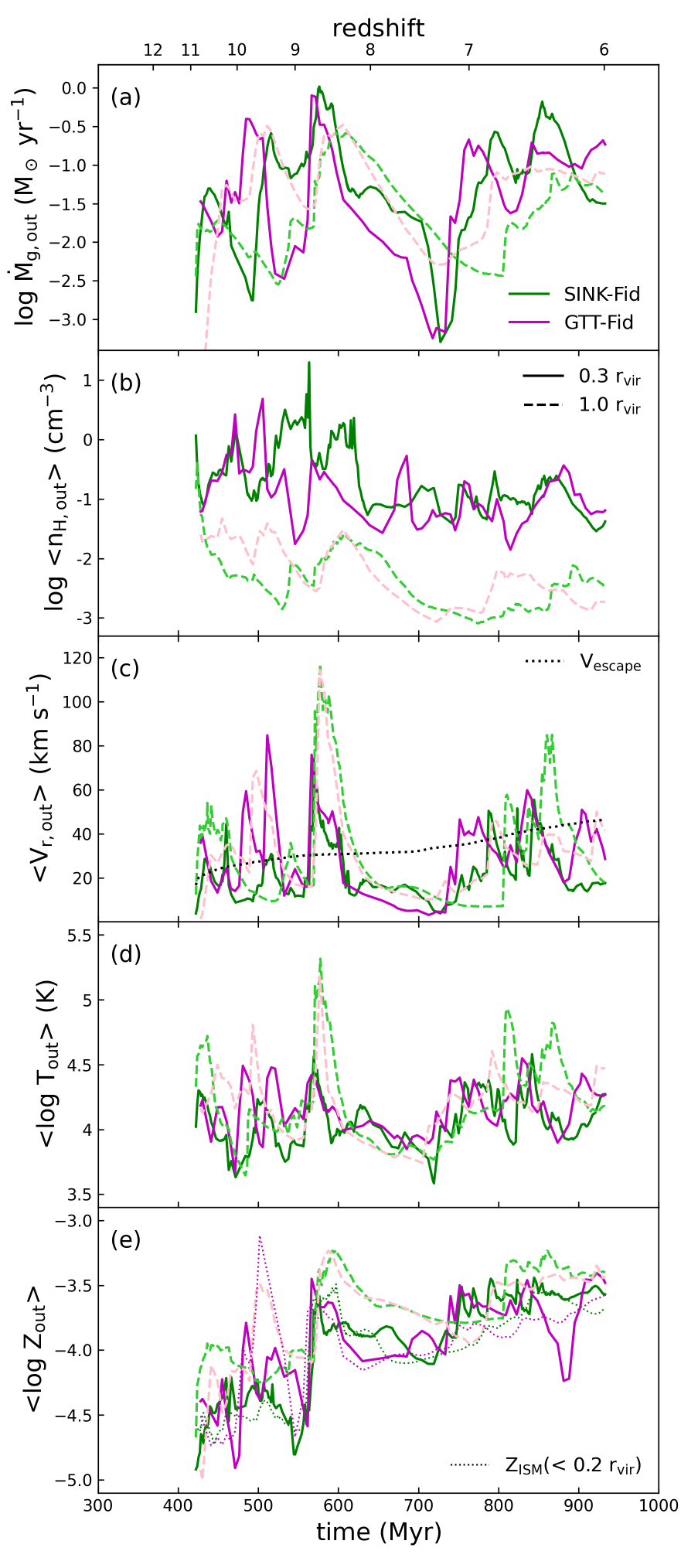}
    \caption{Properties of gas outflows. From top to bottom: (a) the successive panels show outflow rate, (b) hydrogen number density, (c) radial velocity, (d) temperature, and (e) metallicity. Simulations with different star formation models are indicated by different color codes. Solid lines correspond to quantities measured at $r=0.3\, r_{\rm vir}$, while dashed lines are measurements at $1\, r_{\rm vir}$. The black dotted line in panel-(c) indicates the escape velocity from the DMH. Dotted lines in panel-(e) represent mean ISM metallicities within $r=0.2\ r_{\rm vir}$.}
    \label{fig:outflow}
\end{figure}

We now turn to the HI column density distribution ($N_{\rm HI}$), which is another way of measuring the effect of stellar feedback. Most stars form from dense clumps, but these are rapidly dispersed or ionized if feedback is strong. Figure~\ref{fig:violin} shows the distributions of the mean $N_{\rm HI}$ within 50 pc of each star particle, averaged over 6 directions, as a function of stellar age. In the first three bins, where the particles are younger than 1.3 Myr, SN explosions have not yet occurred (although they may have taken place in the neighborhood) and stars in both models are mostly embedded in a dense environment with an average density of $\nH \gtrsim 10^3\, \cmq$. However, the column density gradually decreases with increasing stellar age due to radiation feedback. At $t\gtrsim 10\,{\rm Myr}$, the value of $N_{\rm HI}$ decreases even further due to SN explosions. Interestingly, in \texttt{SINK-Fid} the dense gas is disrupted faster than in \texttt{GTT-Fid} as more clustered and coherent stellar feedback leads to a more dramatic decrease in $N_{\rm HI}$. The column density increases again for $t\gtrsim 30\,{\rm Myr}$, with neutral gas in \texttt{SINK-Fid} recovering more slowly. 

The physical properties of the outflowing gas are shown in Fig.~\ref{fig:outflow}. The outflows were measured on a spherical surface at $0.3\,r_{\rm vir}$ or $1.0\,r_{\rm vir}$, which is uniformly divided into 49,152 pixels (${\rm Nside}=64$) using the \textsc{HEALPix} algorithm \citep{Gorski2005}. We calculated the outflow rate by summing over the pixels with positive radial velocities, as $\dot{M}_{\rm g,out}=\int \rho_{\rm gas} \, v_{\rm rad} \, \Theta(v_{\rm rad}) \, dA_\bot$, where $v_{\rm rad}$ is the radial velocity of the cells, $\Theta$ is the Heaviside step function, and $dA_\perp$ is the area of each pixel with normal vector parallel to the direction of the radial velocity. 
We find that the evolution of outflow rates (or fluxes) generally follows that of the SFRs shown in Fig.~\ref{fig:SFR}, with the outflows being more continuous and of larger amplitude. The flux-weighted outflow rates measured at $6<z<10$ on the inner sphere at $0.3\,r_{\rm vir}$ are 0.40 and 0.25 \msunyr\ in the \texttt{SINK-Fid} and \texttt{GTT-Fid} runs, respectively, indicating that the \texttt{SINK-Fid} galaxy is more effective at generating outflows. In the \texttt{SINK-Fid} galaxy, the flux-weighted outflow measured at $0.3\,r_{\rm vir}$ is denser ($\nH=0.55$ vs. $0.36\,\cmq$) and cooler ($\log T/K=4.11$ vs. $4.23$) than in \texttt{GTT-Fid} (second and fourth panel). This means that the outflows in the \texttt{SINK-Fid} carry more mass overall, although they are more intermittent, because star formation is more bursty. 

Outflows in the inner region of the two galaxies move at velocities comparable to the escape velocity of the DMH (flux-weighted, $\sim30$--$40\,\kms$) (third panel). As such, some fraction of the ejected gas is recycled back into the ISM, rather than expelled from the DMH. Indeed, the outflow rates in the \texttt{SINK-Fid} galaxy are more than halved when measured at the virial sphere (0.14 vs. 0.17 \msunyr) rather than $0.3\,r_{\rm vir}$. 
In contrast, the differences in the properties of the outflowing gas on the virial sphere between the two models are not very pronounced. We attribute this to the fact that there are some random SN explosions from the infalling satellites in the outer halo, which complicate the interpretation at the virial sphere. Nevertheless, the outflow rates measured at both radii are much larger than the SFRs, giving $\left<\dot{M}_{\rm g,out} \right> / \left< \dot{M}_{\rm star}\right>\approx 10$--$25$ in \texttt{SINK-Fid} and \texttt{GTT-Fid}. Note that these mass-loading factors are comparable, albeit slightly lower, to those of other simulated galaxies of similar stellar masses \citep[e.g.,][]{Pandya2021}.

Finally, there is no clear difference in the metallicity of the outflows between the two star formation models (bottom panel of Fig.~\ref{fig:outflow}). The flux-weighted mean metallicities are $\approx 0.01\,Z_{\odot}$ for both runs, which are similar to the metallicities of their ISM (dotted lines in the bottom panel). This is not very surprising, given that the outflows entrain $\sim 50$--$100$ times the mass of the stellar ejecta (20\% of the stellar mass). As a result, the metallicities of the outflows tend to increase with time, following the chemical evolution of the galaxies. However, these turn out to be insensitive to feedback from different star formation models, as the \texttt{SINK-Fid} and \texttt{GTT-Fid} galaxies produce similar amounts of stars (and thus metals) in our simulations.

\section{Discussion}
In this section, we discuss why stars form differently in the runs with two different star formation models. We also assess how the models respond to different input SN energies. Finally, we examine how the burstiness, feedback strength, and density distributions of the star-forming sites change with resolution.

\subsection{Differences in star formation criteria}\label{sec:51}

\begin{figure}
    \includegraphics[width=\linewidth]{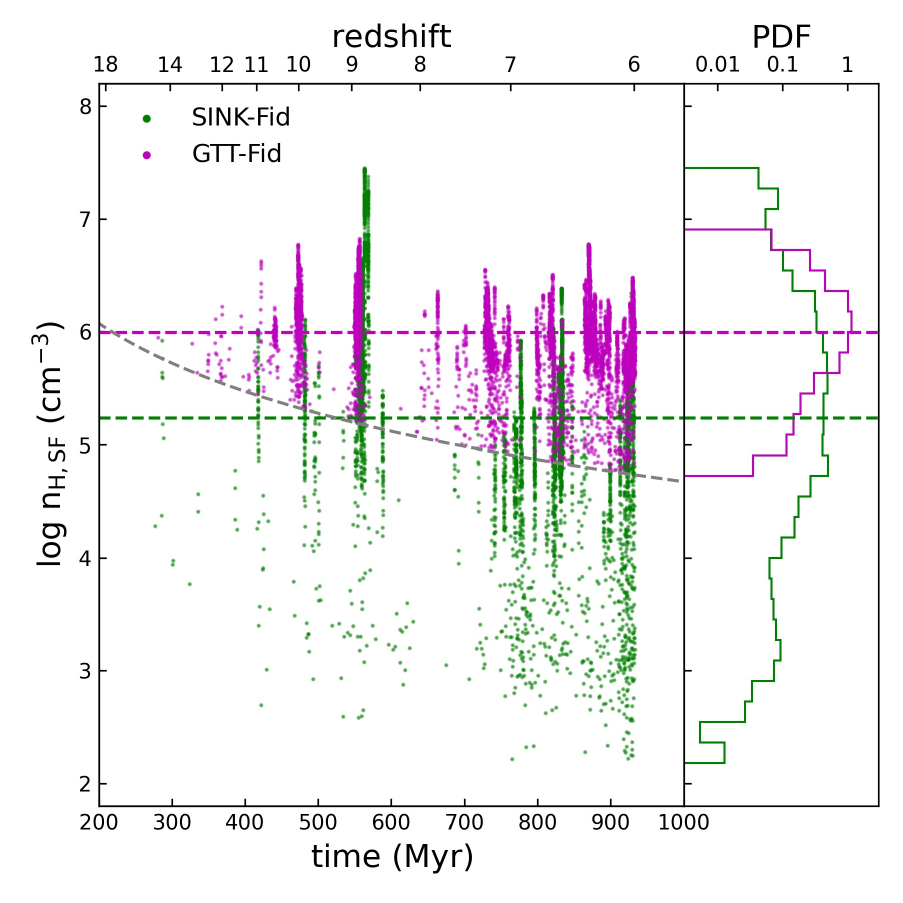}
    \caption{Hydrogen number density of host cells at the time of Pop II star formation ($n_{\rm H,SF}$). Different runs are represented by different color codes, as indicated in the legend. The average density is indicated by an horizontal dashed line. The gray dashed curve corresponds to the minimum density required to create a star particle of mass $500 \,\msun$ in the \texttt{GTT-Fid} without violating the numerical condition that no more than 90\% of a cell gas mass is converted into stars per fine time step.}
    \label{fig:age-density}
\end{figure}

\begin{figure}
    \includegraphics[width=\linewidth]{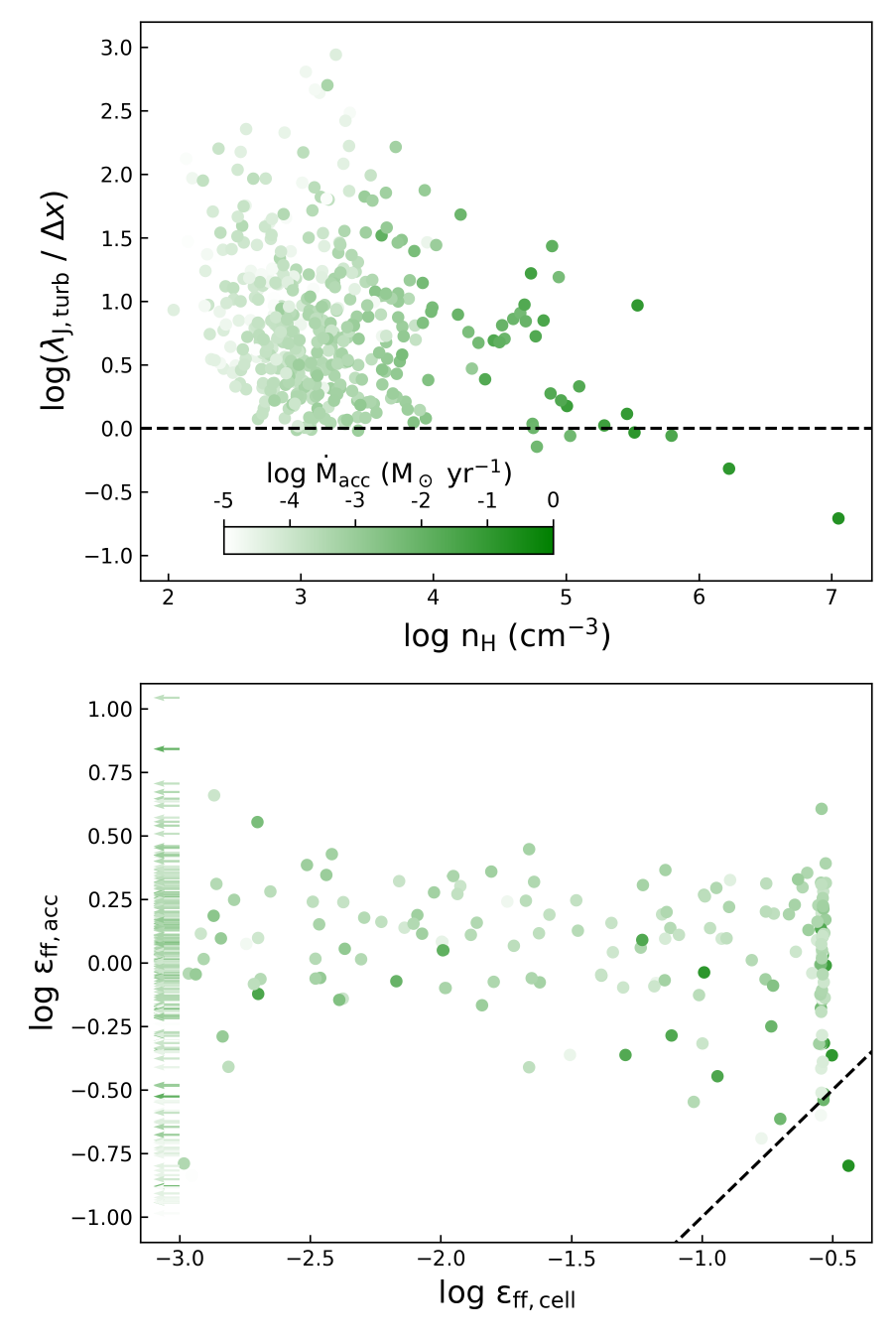}
    \caption{ Properties of efficient accretion events. Top: Turbulent Jeans length over cell size ($\lambda_{\rm J,turb}/\dx$) versus hydrogen number density (\nH) of cells hosting sink particles in \texttt{SINK-Fid}. Note that the sum of gas, stars, sinks, and dark matter particles is used to estimate the total density ($\rho$) of the host cell when calculating $\lambda_{\rm J,turb}$ in Eq.~\eqref{eq:lambda_jt}. We show sink particles with an accretion rate greater than $10^{-5}\, \msunyr$ and \nH\ greater than $100\, \cmq$ from all snapshots. The horizontal dashed line corresponds to $\lambda_{\rm J,turb}=\dx$, below which star formation can be triggered in the \texttt{GTT} runs. 
    Bottom: SFEs per free-fall time, measured directly from the accretion rate and gas mass within the accretion zone ($\varepsilon_{\rm ff,acc}$), versus that derived from the host cell of the sink particle using Eq.~\eqref{eq:eff} ($\varepsilon_{\rm ff,cell}$). Data points with $\varepsilon_{\rm ff,cell} < 10^{-3}\, \msunyr$ are represented by arrows. The black dashed line indicates a one-to-one relationship. The color scale represents $\dot M_{\rm acc}$ in both panels.}
    \label{fig:lambjt+eff}
\end{figure}

In the previous section, we explain how the two star formation models lead to different burstiness. To better understand the differences between these models, we plot in Fig.~\ref{fig:age-density} the hydrogen number density of the cells in which star formation occurs ($n_{\rm H,SF}$). Also shown as a gray line is the minimum gas mass ($556\,\msun$) required in a gas cell to produce a star particle of mass $500\,\msun$ in the \texttt{GTT} runs, as the code imposes a 90\% maximal conversion limit of gas into stars for numerical stability.
We find that star particles in the \texttt{GTT-Fid} run form at densities that are on average a factor of 5.7 higher than in \texttt{SINK-Fid} ($\left< \log n_{\rm H,SF} \right>\approx 5.2$ vs $6.0$). This is partly due to the fact that in \texttt{SINK-Fid} a large amount of gas has already been transferred to the sink particles from accretion over multiple time steps at the time of formation, whereas in \texttt{GTT-Fid} cell gas mass is instantly converted into stars. 
However, even if the sink mass is included in the calculation of $n_{\rm H,SF}$, the average density in \texttt{GTT-Fid} is still larger by a factor of 2.6 than in \texttt{SINK-Fid}. If we restrict our analysis to the period after the strong burst at $z\sim9$ ($6<z<8.5$), where 63 percent of the stars are formed, the difference becomes even more pronounced ($\left< \log n_{\rm H,SF} \right>\approx 4.8$ vs $5.9$). These results suggest that at a given minimal star particle mass, the \texttt{GTT} subgrid model requires more gravitationally bound clouds to form stars than \texttt{SINK-Fid}\footnote{\texttt{GTT-Fid} would produce stars in regions of lower density if a smaller star particle mass is assumed. However, it is important to note that cells with lower masses or densities often fail to satisfy Eq.~\eqref{eq:lambda_jt}. Looking at Fig.~\ref{fig:age-density}, only 30 out of 6,503 pop II stars are formed in cells with mass $\le 600\, \msun$. This suggests that $n_{\rm H,SF}$ in the \texttt{GTT-Fid} model is primarily determined by Eq.~\eqref{eq:lambda_jt} rather than by the choice of the minimum star particle mass.}.

This is further illustrated in the top panel of Fig.~\ref{fig:lambjt+eff} where we plot $\lambda_{\rm J,turb}/\dx$ as a function of gas density for the \texttt{SINK-Fid} run.
Here each point represents a sink particle from different snapshots, color coded by accretion rate. As expected, sink particles accrete actively in dense regions ($\nH\gtrsim 10^4\,\cmq$, upper panel), occasionally forming stars at $\nH\sim 10^6\,\cmq$. Efficient accretion events occur near $\lambda_{\rm J,turb}\sim \dx$, but also in regions where the local turbulent Jeans length is resolved by a few tens of cells. 
This can be attributed to young star particles heating the gas or star-forming clumps exhibiting local turbulence fed by for instance, external accretion and/or galactic fountains. It is also equally possible that turbulent and thermal support in our simulated star-forming clumps may be overestimated due to finite resolution and physical ingredients (see below). Nevertheless, we find that local collapse motion can facilitate accretion onto the sink particle. 
If the turbulent Jeans length criterion ($\lambda_{\rm J,turb}/\dx < 1$) were applied to \texttt{SINK-Fid}, as implemented in \texttt{GTT}, it would result in a significant delay in star formation for these highly accreting sink particles until the gas densities reach a threshold for local gravitational instability.

Conversely, when the accretion rates of potential star-forming sites are measured in the \texttt{GTT-Fid} run (not shown) as done in Sect.~\ref{sec:222}, we see that the high accretion events with $\dot{M}_{\rm acc}\gtrsim 10^{-2}\,\msunyr$ do not necessarily occur in a cell where $\lambda_{\rm J,turb}/\dx<1$, but are mostly distributed over $\lambda_{\rm J,turb}/\dx\lesssim 10$. A possible interpretation is that a cell may be considered locally Jeans unstable if the turbulent Jeans length is resolved by less than $\sim$ 10 cells. Therefore, other subgrid multi-freefall models with a more relaxed Jeans length criterion may share more similarities with our sink particle approach \citep[e.g.,][]{Dubois2021,Girma2024}.

A more important feature of the \texttt{SINK} runs is the predominant formation of star particles from high accretion events. In the \texttt{SINK-Fid} run, $\sim 70\%$ of star particles are formed through accretion rates of $\sim 10^{-3}$--$10^{-1}\,\msunyr$. The high accretion rates results in the formation of star particles with masses of $500\,\msun$ within a short timescale of $\sim0.005$--$0.5\,{\rm Myr}$. We also find that these regions exhibit a more efficient conversion of gas into sinks compared to what is predicted by the multi-freefall approach (bottom panel of Fig.~\ref{fig:lambjt+eff}). Here, $\varepsilon_{\rm ff, acc}\equiv\dot M_{\rm acc}t_{\rm ff}/M_{\rm gas}$ is computed by replacing $dM_\star/dt$ with $\dot M_{\rm acc}$ in Eq.~\eqref{eq:schmidt}, whereas $\varepsilon_{\rm ff, cell}$ is measured for the host cell of sink particles using Eq.~\eqref{eq:eff}. The plot reveals that in regions where sink particles are actively accreting with $\varepsilon_{\rm ff, acc}\sim 1$, $\varepsilon_{\rm ff, cell}$ is estimated to be significantly lower ($\ll 1$), with a maximum value of $\approx 0.5$. Although $\varepsilon_{\rm ff, acc}$ does not precisely coincide with the \eff\ of a turbulent box, these results indicate that more stars are likely to form for a given gas mass when using the sink algorithm compared to the multi-freefall model. This intrinsic difference in \eff\ also suggests that allowing star formation in loosely bound regions does not guarantee a similar level of burstiness in the \texttt{GTT} and \texttt{SINK} runs.

We argue that the elevated accretion efficiency ($\varepsilon_{\rm ff, acc}$) in the \texttt{SINK} run is primarily due to the rapid gravitational collapse of dense star-forming clumps. For the highly accreting events, the accretion rates are mostly calculated using the flux scheme, and we find that the inward radial velocity surrounding sink particles at $\nH\sim10^4\, \cmq$ is $\sim 3$--$5\, \kms$, which is several times higher than the local sound speed. In such environments where dense gas converges, it is difficult to avoid efficient gas collapse\footnote{We have also calculated accretion rates using the Bondi scheme and found that, on average, the accretion rates are only a factor of 2--3 lower than those based on the flux accretion scheme.}. However, it should be noted that the rapid collapse seen in our simulations may be somewhat overestimated due to the absence of magnetic support against gravity. In addition, the turbulence support is also likely to be underestimated at the resolution scale due to the adaptive derefinement or refinement of the computational grid. Accounting for these factors could potentially suppress the gas accretion rate on to sink particles, suggesting that $\varepsilon_{\rm ff, acc}$ in this study may be considered as an upper limit\footnote{We have tested using a GMC simulation that arbitrarily reducing the accretion rate by a factor of two has no dramatic effect on the accreted mass. As the density of the collapsing cloud increases, so does the accretion rate, mitigating the artificial reduction.}.

On the other hand, there is also the possibility of underestimating $\varepsilon_{\rm ff, cell}$ in the \texttt{GTT} model. To determine $\varepsilon_{\rm ff, cell}$ using Eq.~\eqref{eq:eff}, the local Mach number and $\alpha_{\rm vir}$ must be known. 
Since the \texttt{GTT} model does not explicitly model turbulence \citep[e.g.,][]{Kretschmer2020}, we derive the velocity dispersion on resolution scale ($\Delta x_{\rm min}$) using kinematic information from six local neighbors. However, since the turbulence strength is scale-dependent \citep[$\sigma_{l} \propto l^{0.5}$, e.g.,][]{Larson1981}, turbulence measured from immediate neighbors ($l\sim 3\,\Delta x_{\rm min}$) may be overestimated by a factor of $\sim 1.7$ (or $\sim 3$ for $\alpha_{\rm vir}$ for supersonic turbulence). Furthermore, our sound speed is probably overestimated due to the lack of low-temperature coolants such as dust, which often results in a virial parameter of $\alpha_{\rm vir}\gtrsim10$. The $\lambda_{\rm J,turb}<\dx$ condition would then inhibit star formation in these environments (see Fig.~\ref{fig:eff_map}). Even if the condition is relaxed, the use of Eq.~\eqref{eq:eff} may still be questionable, since many of the star formation events in the \texttt{GTT} runs occur in transonic regimes ($\mathcal{M}\lesssim 2$) where shock-induced multi-freefall models may not apply \citep[see Sect. 2.5 of][for discussion]{FK12}. In short, simulations geared toward better capturing the cosmic turbulence cascade are needed to better estimate \eff\ in different collapse environments and thus improve the accuracy of the GTT scheme.

\begin{figure}
    \includegraphics[width=\linewidth]{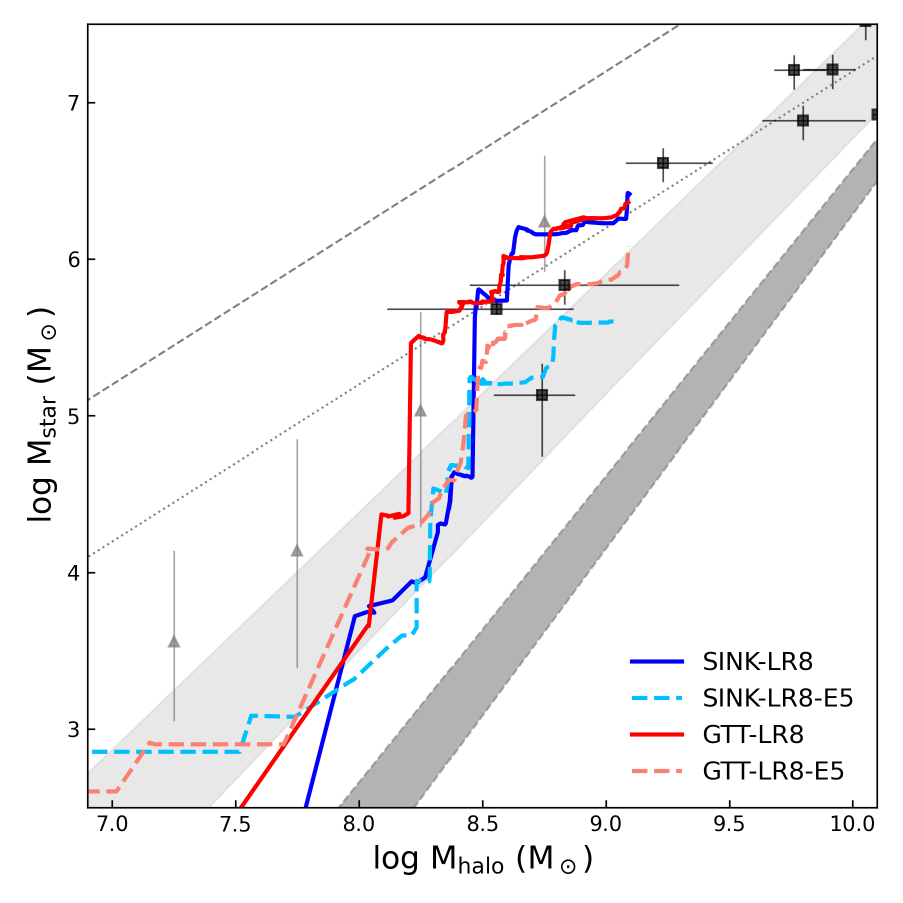}
    \caption{Stellar mass-to-halo mass relation for four different runs with low resolution ($\dxmin=5.5\,{\rm pc}$ at $z=6$, \texttt{LR8}). Dashed lines indicate the runs with SN energy five times stronger, $E_{\rm SN}=5\times10^{51}\, \rm erg$ (\texttt{E5}). Other gray lines and data points are the same as in Fig.~\ref{fig:SMHM}. The stellar mass is reduced by a factor of $\sim6$ and $\sim2$ in the \texttt{SINK-LR8} and \texttt{GTT-LR8} runs, respectively.}
    \label{fig:E5_SMHM}
\end{figure}

\subsection{Effect of supernova energy}
In most of our simulations with resolutions between 0.7 and 11 pc, the galactic stellar masses in the two star formation models do not differ by more than a factor of 2, despite the fact that star formation is more bursty and the final momentum from SNe is larger in the \texttt{SINK} runs. This suggests that galaxy growth in this dwarf-sized DMH regime is well regulated by SN feedback regardless of the star formation model \citep{Xu2016,Kimm2017}. However, it is possible that the simulations underestimate the impact of SNe, because we do not consider additional pressure from cosmic rays \citep{Rodriguez-Montero2022}, or top-heavy initial mass functions \citep[e.g.,][]{Katz2022IMF,Cameron2024}, or hypernovae \citep[e.g.,][]{Kobayashi2006, Bhagwat2024}. Motivated by these caveats, and to confirm that our main conclusion about burstiness remains the same with different feedback strengths, we re-run the \texttt{GTT-LR8} and \texttt{SINK-LR8} simulations with an enhanced SN explosion energy of $E_{\rm SN}=5\times10^{51}\, \rm erg$. We will call these runs \texttt{E5}, as opposed to the \texttt{E1} runs with $E_{\rm SN}=10^{51}\,{\rm erg}$. To reduce the computational cost, we adopt a maximum refinement which is three levels lower (\texttt{LR8}), corresponding to $\dxmin=5.5\, \rm pc$ at $z=6$, but still yields reasonably similar burstiness and stellar masses (see next section).

\begin{figure}
    \includegraphics[width=\linewidth]{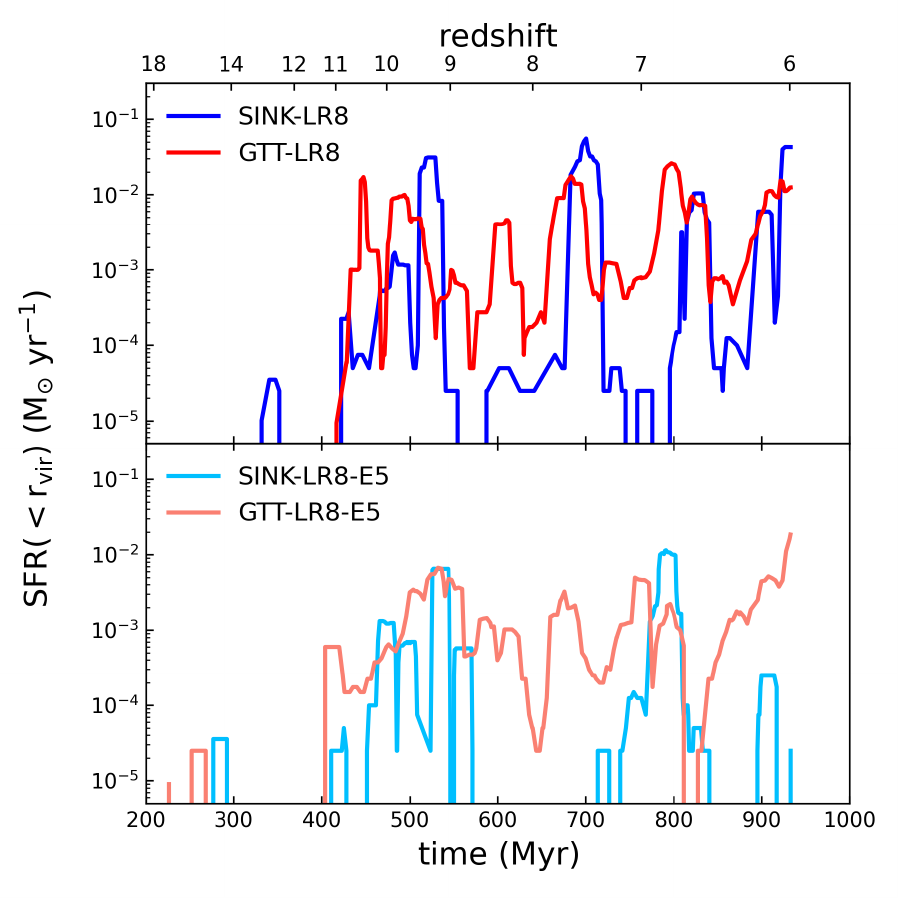}
    \caption{Star formation rates in the \texttt{E1} (top) and \texttt{E5} runs (bottom). SFR are averaged over a 20 Myr timescale. We find that star formation is still more bursty in the SINK run than in the \texttt{GTT} run, even when the SN energy is artificially increased.}
    \label{fig:E5_SFR}
\end{figure}

Figure~\ref{fig:E5_SMHM} shows the growth of galactic stellar mass down to $z=6$ in the \texttt{LR8} simulations. Although star formation in the early phase is slightly different, the two \texttt{E1} halos with the \texttt{GTT} and \text{SINK} models produce a similar amount of stellar mass by $z=6$. As in the fiducial runs, the \texttt{SINK-LR8-E1} simulation produces greater number of stars over short periods of time compared to its \texttt{GTT} counterpart, as demonstrated by the more pronounced and concentrated peaks in its star formation history (see top panel of Fig.~\ref{fig:E5_SFR}). When the SN energy is increased by a factor of 5, both models produce fewer stars due to stronger explosions than in the \texttt{E1} runs, while maintaining the same trend in burstiness (the bottom panel of Fig.~\ref{fig:E5_SFR}). However, the exact response to increased SN energy in the final stellar mass is different. In the \texttt{GTT} model, stars continue to form between the two mergers at $550\,{\rm Myr} < t < 750 \,{\rm Myr}$, while the \texttt{SINK} run completely suppresses star formation for $\sim 150\, \rm Myr$. In addition, the stellar mass formed during the first merger at $z\sim 9$ decreases to a level similar to that of the \texttt{GTT} run. Consequently, the final stellar mass in \texttt{SINK-LR8-E5} is further reduced by a factor of three compared to \texttt{GTT-LR8-E5}.

\begin{figure}
    \includegraphics[width=\linewidth]{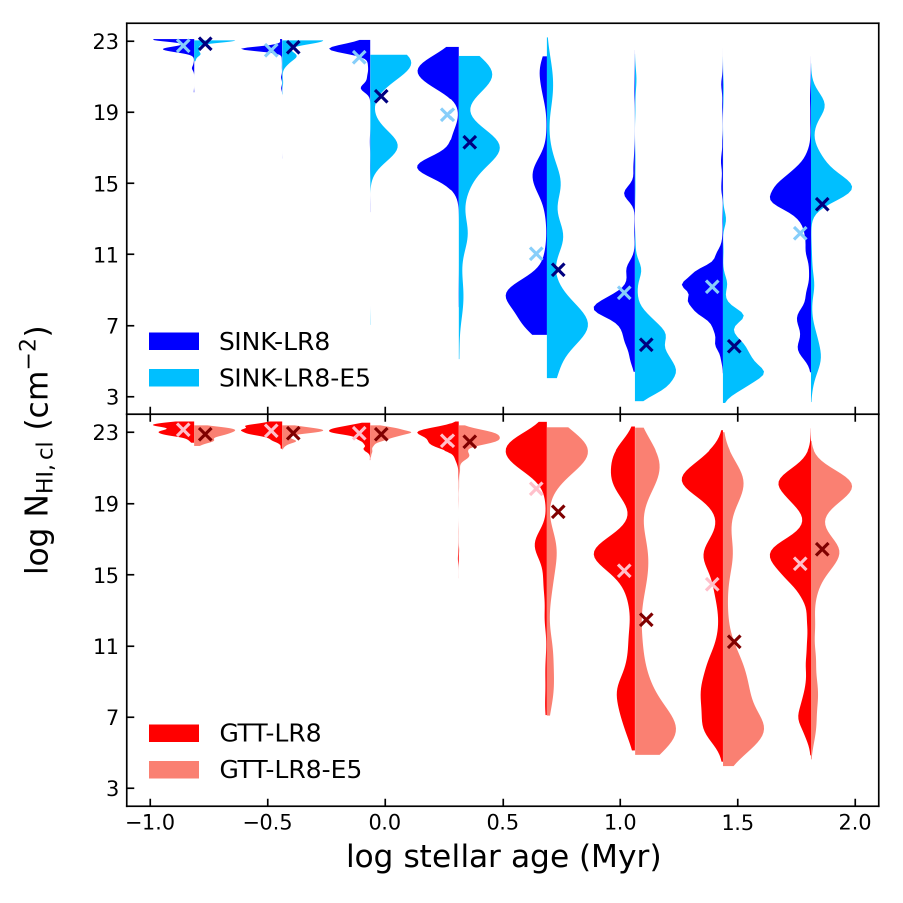}
    \caption{Same as Fig.~\ref{fig:violin}, but for the \texttt{SINK-LR8} (top) and \texttt{GTT-LR8} runs (bottom). The mean column density of neutral hydrogen within 50 pc of each young star particle ($N_{\rm HI,cl}$) decreases faster in the \texttt{SINK} runs than in the \texttt{GTT} runs for both \texttt{E1} and \texttt{E5} simulations.}
    \label{fig:E5_violin}
\end{figure}

The impact of strong feedback can also be seen in the HI column density on cloud scales (50 pc) measured around each star particle ($N_{\rm HI,cl}$, Fig.~\ref{fig:E5_violin}). Consistent with the fiducial runs, the \texttt{SINK-LR8-E1} run is more effective at dispersing dense gas clouds, and the distribution of $N_{\rm HI,cl}$ is skewed to lower values than in \texttt{GTT-LR8-E1}. More importantly, when the feedback strength is boosted (\texttt{E5}), the typical $N_{\rm HI,cl}$ becomes lower for stars with ages greater than $\sim 5$--$10\,{\rm Myr}$ in the \texttt{GTT} runs. In contrast, $N_{\rm HI,cl}$ is already reduced around stars younger than 1 Myr in the \texttt{SINK-LR8-E5} run. This may seem counter-intuitive, given that the minimum lifetime for SN progenitors is much larger ($\approx 4\,{\rm Myr}$). However, stars form in clusters, with stars formed first pre-reducing the column density around the others. This may also have happened in \texttt{SINK-LR8-E1}, although the explosions may not have been powerful enough to disperse dense clouds in that case. Figure~\ref{fig:E5_violin} clearly suggests that bursty star formation via the sink particle algorithm may become increasingly efficient at disrupting GMCs and driving powerful winds when stellar feedback is intrinsically stronger.

\begin{figure}
    \includegraphics[width=\linewidth]{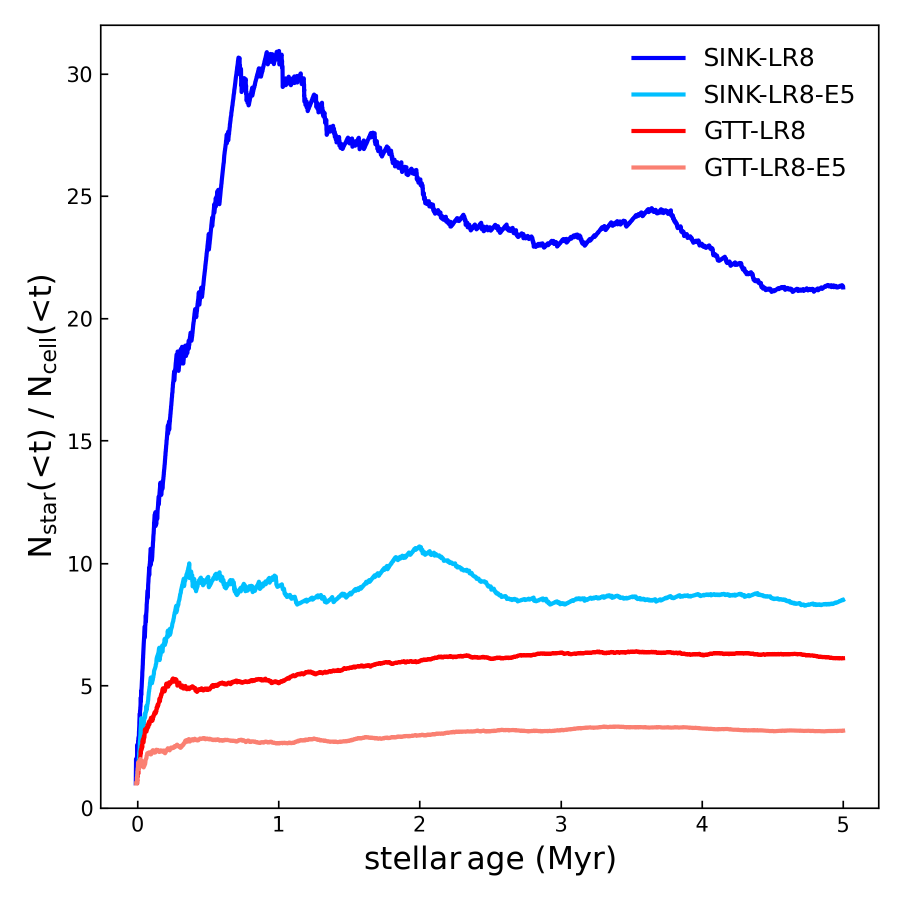}
    \caption{Ratio between the number of star particles younger than a certain age ($N_{\rm star}(<t)=M_{\rm star}(<t)/m_{*,\rm min}$) and the number of cells containing at least one young star particle with age ($N_{\rm cell}(<t)$). We include stars inside the main halo in all snapshots. Simulations with different star formation models and SN energies are shown with different color codes, as indicated in the legend.}
    \label{fig:E5_clustering}
\end{figure}

Figure~\ref{fig:E5_clustering} also demonstrates that stars form in a more clustered fashion in the \texttt{SINK} runs than in \texttt{GTT}. We measure the ratio between the cumulative number of young stars ($N_{\rm star}(<t)\equiv M_{\rm star}(<t)/m_{*,\rm min}$) and the number of cells containing at least one star particle ($N_{\rm cell}(<t)$), as a simple proxy for the degree of clustering, as a function of stellar age. Note that this ratio increases with higher clustering, and a ratio of 1 means one star particle with $m_{*,\rm min}$ per star-forming cell. We find that the typical number of young stars per cell is about five times larger in the \texttt{SINK} runs than \texttt{GTT}. This is because the sink particles can accrete very efficiently and form multiple star particles in dense regions, whereas the turbulent Jeans length condition and the low SFE of the \texttt{GTT} model tend to prevent a single cell from forming a large number of stars. 
For example, the typical velocity of the sink particles we measure from the \texttt{LR8} simulation is $8$--$10 \,\kms$, and it thus takes $\sim 0.5\,{\rm Myr}$ to move out a cell of size $\dx=5.5\,{\rm pc}$. If the accretion rate remains high ($\dot{M}_{\rm acc}\sim10^{-2}\,\msunyr$), it could easily yield $5,000\,\msun$ of stars (or 10 star particles with $m_{*,\rm min}$) formed in the cell during that time. 
Of course, the exact amount of stellar mass per cell depends on the SN feedback strength as well as grid resolution, as briefly discussed in Sect.~\ref{sec:32} and as can be seen in Fig.~\ref{fig:E5_clustering}. However, the comparison between the ratios $N_{\rm star}/N_{\rm cell}$ from simulations with different feedback energies and resolutions suggests that stars form more coherently in space and time in the \texttt{SINK} runs. Thus, we argue that more clustered star formation, via collective feedback processes, may help prevent the galactic center from continuously forming stars over long periods of time and produce overly massive bulges \citep[e.g.,][]{Joung2009}.

\subsection{Resolution effect}
Turbulent GMC simulations show that a spatial resolution of $\sim 0.25\,{\rm pc}$ is necessary to achieve a reasonable convergence in star formation efficiency (see Sect.~\ref{sec:A2}). However, running cosmological simulations at such a high resolution is prohibitively expensive, even with the zoom-in technique. We therefore had to settle for a lower resolution of 0.7 pc (0.4 pc) at $z=6$ ($z=11$) in our fiducial run. It is thus important to assess the extent to which our main results have converged.

\begin{figure}
    \includegraphics[width=\linewidth]{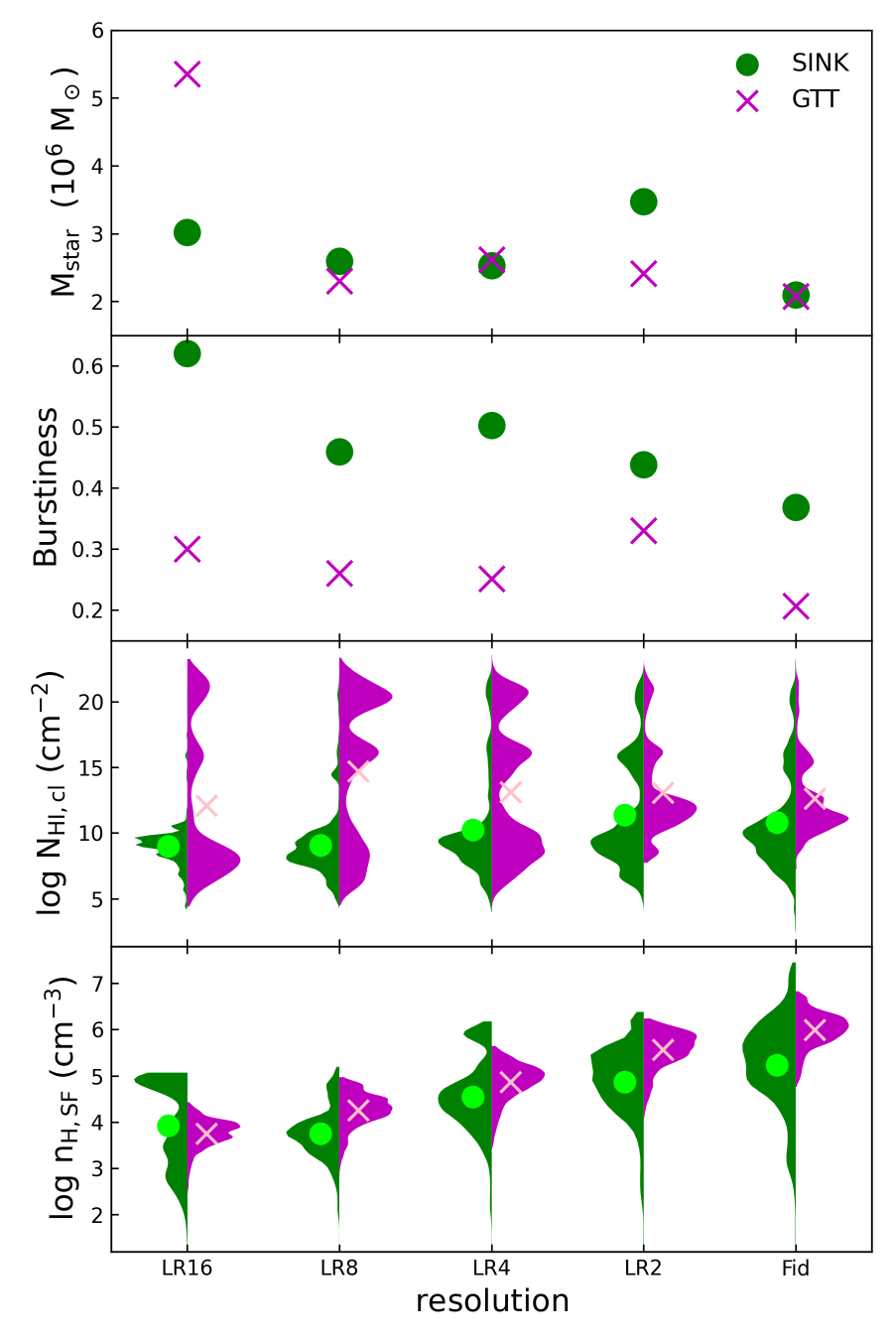}
    \caption{Resolution effects on four quantities related to star formation. From top to bottom, we plot final stellar mass at $z=6$, burstiness parameter $B$, column density of neutral hydrogen around stars of age $7.5 \le t \le 42\,{\rm Myr}$ (i.e., the sum of the sixth and seventh bins of Fig.~\ref{fig:violin}), and hydrogen number density of star-forming cells. Filled circles and crosses in the two bottom panels represent mean logarithmic densities.}
    \label{fig:res_effect}
\end{figure}

Figure~\ref{fig:res_effect} illustrates that while quantitatively convergence has not yet been reached, our key findings exhibit the same trend regardless of resolution. First, the predicted stellar masses within the main halo at $z=6$ are reasonably similar (in general better than a factor of two (top panel), except for the lowest resolution run (\texttt{GTT-LR16}, $\dxmin=11\,{\rm pc}$)), suggesting that star formation is self-regulating. This is consistent with \citet{Kim2017} where the star formation rates in the ISM patch simulations begin to converge with the sink particle algorithm at $\dx \le 16\, \rm pc$.

Second, the burstiness parameter (Eq.~\eqref{eq:B-param}) is always higher in the \texttt{SINK} runs than in the \texttt{GTT} runs (second panel). We note that burstiness tends to decrease with increasing resolution as gas is able to collapse further and its subsequent disruption by stellar feedback occurs in multiple clumps of smaller mass. \citet{Faucher2018} also showed analytically that the SFR variability increases with decreasing number of gravitationally bound clouds. One might expect the burstiness parameter to decrease more rapidly with increasing resolution in the \texttt{SINK} runs, leading to a convergence of $B$ between the two star formation models. Although not shown here, we also run the \texttt{SINK} run at a higher resolution ($\dxmin=0.35\, \rm pc$ at $z=6$) down to $z=8.5$ and found that the burstiness parameter is almost identical ($B=0.351$) to that of \texttt{SINK-Fid} measured at the same redshift range ($B=0.353$), suggesting that there is an intrinsic difference in burstiness between the two star formation models. Accordingly, UV luminosities of the \texttt{SINK} galaxies generally vary more dramatically and are therefore, for a given stellar mass, often larger than those of their \texttt{GTT} counterparts.

The third panel of Fig.~\ref{fig:res_effect} shows the column density distribution of neutral hydrogen measured within 50 pc from each star particle. Again, the typical column densities around young stars with ages of $7.5 \le t/{\rm Myr} \le 42$, corresponding to the sixth and seventh bins in Fig.~\ref{fig:violin}, are lower in the \texttt{SINK} runs. Together with the burstiness parameter from the second panel, we conclude that bursty star formation clears local star-forming clumps earlier, and that the subsequent expansion of SN remnants is thus less hindered by the prevailing ISM conditions. 

Finally, the last panel displays the hydrogen number density distribution within grid cells where star particles form ($n_{\rm H,SF}$). Unsurprisingly, the typical density of star-forming cells increases with increasing resolution, because, as previously mentioned, the gravitational collapse of star-forming clouds can be tracked further. Nevertheless, our conclusion that star particles are born in lower density environments in the \texttt{SINK} runs than in \texttt{GTT} still holds at most resolutions. Again, the exception is the lower resolution \texttt{SINK-LR16} run, which shows a pronounced tail of high density star-forming cells ($n_{\rm H,SF}\sim 10^5\,\cmq$). This occurs during the first merger-triggered starburst at $z\sim 9$: a single event with SFR close to $0.1\, \msunyr$ which accounts for the build-up of more than 50\% of the total stellar mass at $z\sim 6$. This also significantly increases the burstiness (second panel), but is very likely a stochastic rather than a systemic outcome.

Taken together, we conclude that the self-regulating nature of star formation and feedback in our runs allows for a reasonable amount of convergence over a wide range of resolutions. However, we caution that this may turn out to be a reflection of our simulated halo being less prone to overcooling. For instance, using a simple density-based star formation model with a mechanical SN feedback scheme, \citet{Smith2019} showed that star formation is only regulated in one of the five dwarf galaxies they simulate. In this respect, further investigation of more (massive) systems is needed, which we intend to pursue in a forthcoming paper.

\begin{figure*}
    \includegraphics[width=\linewidth]{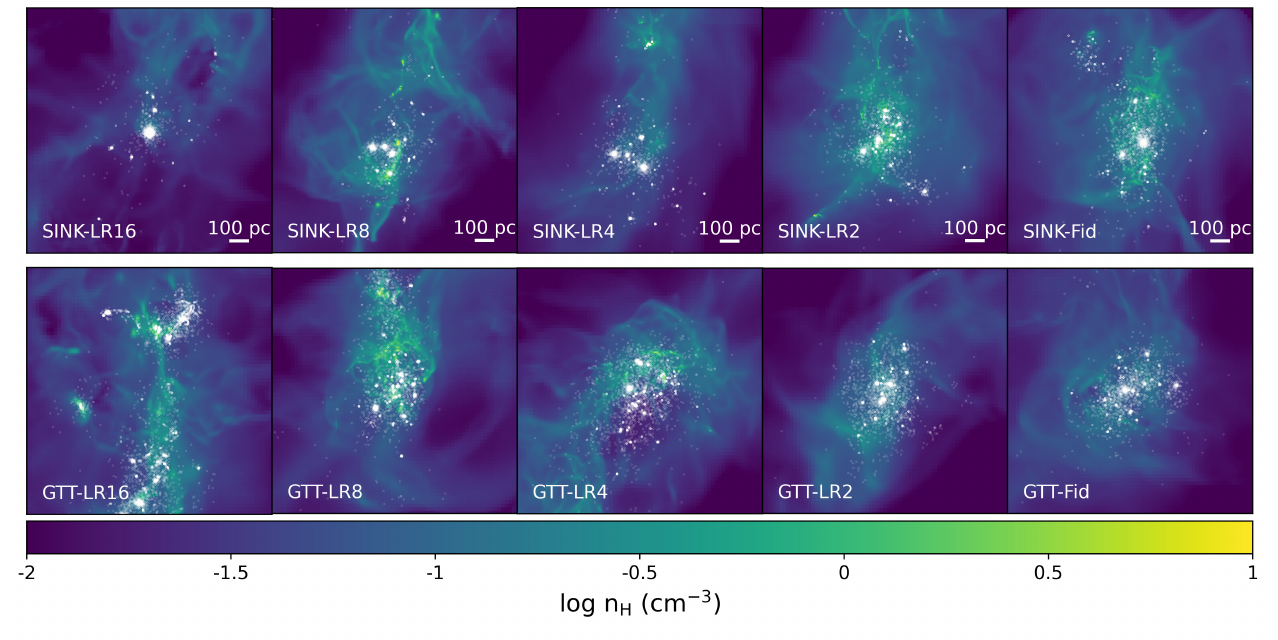}
    \caption{Projected images of hydrogen number density around the galactic center at $z=6$ in simulations with different resolutions. From left to right the maximum resolution increases from 11 to 0.7 pc (physical). Each panel is $0.3\,r_{\rm rvir}$ on a side. The top (bottom) panels show ISM and stellar distributions in the \texttt{SINK} (\texttt{GTT}) runs. Star particles are represented as white dots.}
    \label{fig:res_map}
\end{figure*}

\section{Summary}
In this study, we investigate the impact of star formation on the growth of a dwarf galaxy hosted by a DMH with mass of $10^9\,\msun$ at $z=6$. We  compare a subgrid model based on a local gravo-thermo-turbulent (\texttt{GTT}) condition \citep{Kimm2017} and a more direct approach, which evaluates gas accretion onto sink particles, based on the algorithm developed by \citet{Bleuler2014} and \citet{Bleuler2015} (\texttt{SINK}). Performing a series of simple numerical experiments with \textsc{ramses} \citep{Teyssier2002}, we calibrated the parameters required by the sink particle algorithm (see Sect.~\ref{sec:appen}) and then applied this latter to cosmological radiation-hydrodynamic simulations. Our main results can be summarized as follows:

\begin{itemize}
    \item In cosmological zoom-in simulations of a dwarf galaxy, star formation histories in the \texttt{SINK} model are more bursty than in the \texttt{GTT} model, at a wide range of resolutions ($0.7 < \dxmin < 11 \,{\rm pc}$). The main reason for this behavior is rapid gas accretion onto sinks in collapsing star-forming clumps, which results in \eff\ much higher than that predicted by the multi-freefall model. Furthermore, at moderate densities, where the turbulent Jeans length is locally resolved, star formation is not allowed in the \texttt{GTT} subgrid model, whilst sinks continue to accrete at significant rates and form stars. (Fig.~\ref{fig:lambjt+eff}).
    
    \item The bursty nature of the \texttt{SINK} model makes SN explosions more coherent in space and time (Fig.~\ref{fig:E5_clustering}), thus increasing feedback impact for a given amount of stars formed (Figs.~\ref{fig:SN}, \ref{fig:violin}, \ref{fig:E5_violin}). As a result, the number of SNe exploding at $\nH\gtrsim10^3\,\cmq$ is dramatically reduced in the \texttt{SINK} runs. Due to more coherent feedback, the \texttt{SINK} runs disrupt gas clouds more efficiently (Figs.~\ref{fig:violin}, \ref{fig:res_effect}), thus generating denser and stronger outflows (Fig.~\ref{fig:outflow}). 
    
    \item Whilst the final galaxy stellar mass at $z=6$ is similar in both models (Fig.~\ref{fig:SMHM}), star formation in \texttt{GTT} requires gravitationally bound clumps, with SN explosions generally occurring in high density regions and leading to more spread out star formation. By contrast, the \texttt{SINK} model allows a large number of stars to form in a spatially clustered way over a short period of time (Figs.~\ref{fig:SFR}, \ref{fig:E5_SFR}).

    \item When the amount of energy released by SN is artificially increased by a factor 5 ($E_{\rm SN}=5\times10^{51}\, \rm erg$), the bursty nature of the \text{SINK} model becomes more pronounced (Fig.~\ref{fig:E5_SFR}). The combined effect of clustered star formation and stronger individual SN explosion disperses star-forming clumps more easily, and star formation is thus more efficiently regulated in the \texttt{SINK} model. This results in a galaxy stellar mass lower by a factor 3 in the \texttt{SINK-LR8-E5} run than in its \texttt{GTT-LR8-E5} counterpart.
\end{itemize}

Using numerical experiments, we have demonstrated that modeling star formation with a sink particle algorithm, a more direct method than multi-freefall-based models, increases burstiness and, as a result, helps alleviate the overcooling problem which has crippled (and still does) galaxy formation simulations for decades \citep[e.g.,][]{Joung2009}. In a separate attempt, Han et al. ({\sl in prep}) have also performed RHD simulations of an idealized disk with properties mimicking NGC 300 and find that the disruption of GMCs is considerably more efficient in their run using the sink algorithm as well. However, we draw the attention of the reader to the fact that only one halo has been studied in the work we present here and therefore that further investigation of more (massive) halos is needed, as it has been well established that star formation variability depends on galaxy mass and gas fraction \citep{Faucher2018}. Finally, although we have covered as wide a range of resolutions as possible and found reasonable convergence in stellar mass, we caution this is not the case both for ISM structure and stellar distribution, as can be seen in Fig.~\ref{fig:res_map}. Future simulations are needed to better understand the formation and evolution of ISM structures on (sub)sonic scales by resolving gas fragmentation when cooled by different metallic species, molecules, and dust \citep[e.g.,][]{Katz2022Prism} in a magnetized medium, which ultimately leads to the formation of individual stars.

\begin{acknowledgements}
We thank the referee for the constructive comments that improved the quality of the manuscript. TK is supported by the National Research Foundation of Korea (NRF) grant funded by the Korea government (MSIT) (2022R1A6A1A03053472 and 2022M3K3A1093827), and acted as the corresponding author. CK and DH are partly supported by the National Research Foundation of Korea (NRF) grant funded by the Korea government (2020R1C1C1007079). This work was supported by the Yonsei Fellowship, funded by Lee Youn Jae. The supercomputing time for numerical simulations was kindly provided by KISTI (KSC-2023-CRE-0162), and large data transfer was supported by KREONET, which is managed and operated by KISTI. This work was also performed using the DiRAC Data Intensive service at Leicester, operated by the University of Leicester IT Services, which forms part of the STFC DiRAC HPC Facility (www.dirac.ac.uk). The equipment was funded by BEIS capital funding via STFC capital grants ST/K000373/1 and ST/R002363/1 and STFC DiRAC Operations grant ST/R001014/1. DiRAC is part of the National e-Infrastructure.
\end{acknowledgements}

\bibliographystyle{aa}
\bibliography{references}


\appendix
\section{Testing the SINK model}\label{sec:appen}
To determine the parameters for the sink particle algorithm that can most accurately model the gravitational collapse, we perform two different types of idealized simulations using an AMR code \textsc{ramses} \citep{Teyssier2002}. We first simulate a series of singular isothermal spheres, changing the accretion scheme and the accretion radius. We then run turbulent GMC simulations using the result from these former simulations to study the effect of resolution on stellar mass.

\subsection{Gravitational collapse of singular isothermal spheres}\label{sec:A1}

\subsubsection{Simulation setup}
Assuming an initially static singular isothermal sphere, the density and radial velocity profiles at $t=0$ can be written as
\begin{equation}
    \rho_{\rm SIS}(r,\,t=0)=\frac{A c^2_{\rm s}}{4\pi G r^2}, \quad u_{\rm SIS}(r,\,t=0)=0,
\end{equation}
where $A$ is the overdensity constant. The sphere is in unstable hydrostatic equilibrium when $A=2$. When $A>2$, the sphere collapses globally because gravity is stronger than the thermal pressure gradient everywhere inside the sphere. As it collapses, the density and velocity profiles change with time, but the central accretion rate $\dot{M} (r\rightarrow 0)$ is a constant for a given $A$ \citep{Shu1977}. This property allows us to test the accuracy of different accretion schemes.

We use \textsc{ramses} \citep{Teyssier2002} to compare the mass growth of a sink particle with the analytic solution given by \citet{Shu1977}. We place a sink particle at the center of the sphere in a $(32\, \rm pc)^3$ box, with three different $A$ (2.2, 4, 90) and three different accretion radii ($\dxmin, 2\dxmin, 4\dxmin$). The different values of $A$ represent gas clouds with initial average densities of $\nH=10$--$103\,\cmq$ or gas mass of $36$--$374\,\msun$ within a radius of 3 pc. The simulations are carried out with two different accretion schemes: Bondi-Hoyle accretion and flux accretion\footnote{We do not test the accretion scheme based on a density threshold because the mean gas density inside the accretion zone is already less than $\rho_{\rm th}= \rho_{\rm LP}(0.5\dxmin)$ for $A=2.2$, $4$ and thus the accretion rate in such a scheme would be zero by construction, which is incorrect.}. The gas temperature is set to 10 K. 

The initial density and velocity profiles are set to follow the analytic solution at $t=9\, \rm Myr$ \citep{Shu1977}. The epoch is chosen arbitrarily, and the initial sink mass is set accordingly to $\dot{M}_{\rm ana}\, t$, where $\dot{M}_{\rm ana}$ is the analytic accretion rate, which depends only on $A$. The cells within a radius of 6.4 pc are maximally refined with $0.0625\, \rm pc$, and the profiles are truncated at $r=6\, \rm pc$. The gas density outside the sphere is set to $\rho_{\rm SIS}(r=6\, {\rm pc},\, t=9 \,{\rm Myr})$, while its radial velocity is set to $u_{\rm SIS}(r=6\, {\rm pc},\, t=9\, {\rm Myr})$. We stop the simulations at $t_{\rm end}=M(r<3\, {\rm pc})/\dot{M}_{\rm ana}$, so that the accretion at the center is not affected by the gas outside the sphere. We note that this timescale differs from the free-fall time in the sense that while the free-fall time of a uniform cloud is independent of radius, $t_{\rm end}$ of an isothermal sphere increases with size, since the central accretion rate is fixed.

\begin{figure}
    \includegraphics[width=\linewidth]{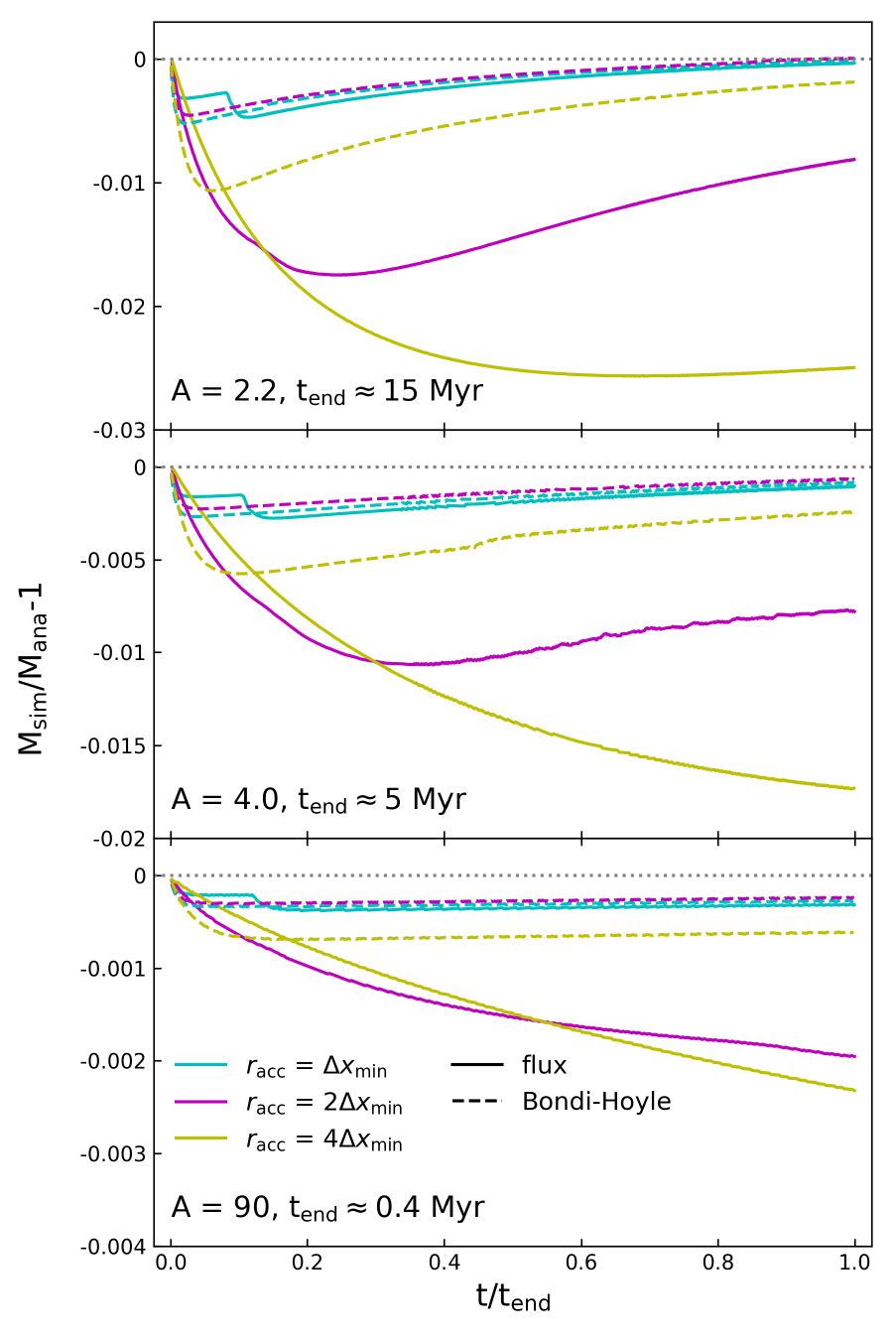}
    \caption{Relative error of the sink particle mass in collapsing isothermal sphere simulations with two different accretion schemes. We examine two values of $A$ (2.2, 4.0) in Table 1 of \citet{Shu1977} and an extreme case (90) where the mean gas density inside the accretion zone is similar to $\rho_{\rm th}$. The lines with different colors represent the results of runs with different accretion radii, as indicated in the legend. The flux and Bondi-Hoyle accretion runs are shown as solid and dashed lines, respectively.}
    \label{fig:SIS_scheme}
\end{figure}

\subsubsection{Determining the accretion method}
Figure~\ref{fig:SIS_scheme} shows relative errors on the sink mass in the collapsing isothermal sphere simulation runs with two different accretion schemes, compared to the analytic solution of \citet{Shu1977}. From top to bottom, the relative errors are computed for different values of $A$, while the different colors and line styles correspond to different sizes for the accretion zone and the accretion schemes. 
Several interesting features appear in this plot. First, the analytic accretion rates are well reproduced in all cases with different $A$ and \racc. Among our 18 test simulations, the one with the largest error is 97\% accurate at the end of the simulation. Second, the computed accretion rates are underestimated compared to the true values. This happens because the analytic solution for the central accretion rate is determined by the asymptotic behavior at $r\rightarrow 0$, while the radius at which the mass flux is computed is larger. 
For a flux accretion scheme, the accretion rate must be constant at any radius if the product of the density and velocity profiles is proportional to $r^{-2}$. This is true for the very inner part of the sphere, where the density and velocity profiles can be approximated as $\rho \propto r^{-1.5},\, u\propto r^{-0.5}$. However, this scaling breaks down at larger radii, where the slope of the two profiles changes to $\rho \propto r^{-2},\, u\propto r^{-1}$. Therefore, as \racc\ becomes larger, the product of the two analytic profiles diverges slightly from $r^{-2}$, leading to an underestimation of the accretion rate. 
For the Bondi-Hoyle accretion scheme, where the accretion rate is proportional to both the kernel-weighted mean density within \racc\ \citep{Krumholz2004} and the sink plus gas mass, the trend is somewhat different, but the accretion rates are still slightly underestimated. In this scheme, as \racc\ becomes larger, the mean gas density decreases but the total mass within \racc\ increases, and we find that the two effects cancel out when $\racc=1$--2$\dxmin$, while the accreted mass becomes less accurate with $\racc=4\dxmin$. 

These experiments suggest that a smaller \racc\ is better for accurately modeling accretion rates. However, \citet{Bleuler2014} pointed out that too small accretion radii may not be able to represent the physical properties of the gas flow, especially if local gas structures are underresolved. Following their argument, we adopt $\racc=2\dxmin$ as a compromise between accuracy and the ability to account for the local environment.

\begin{figure}
    \includegraphics[width=\linewidth]{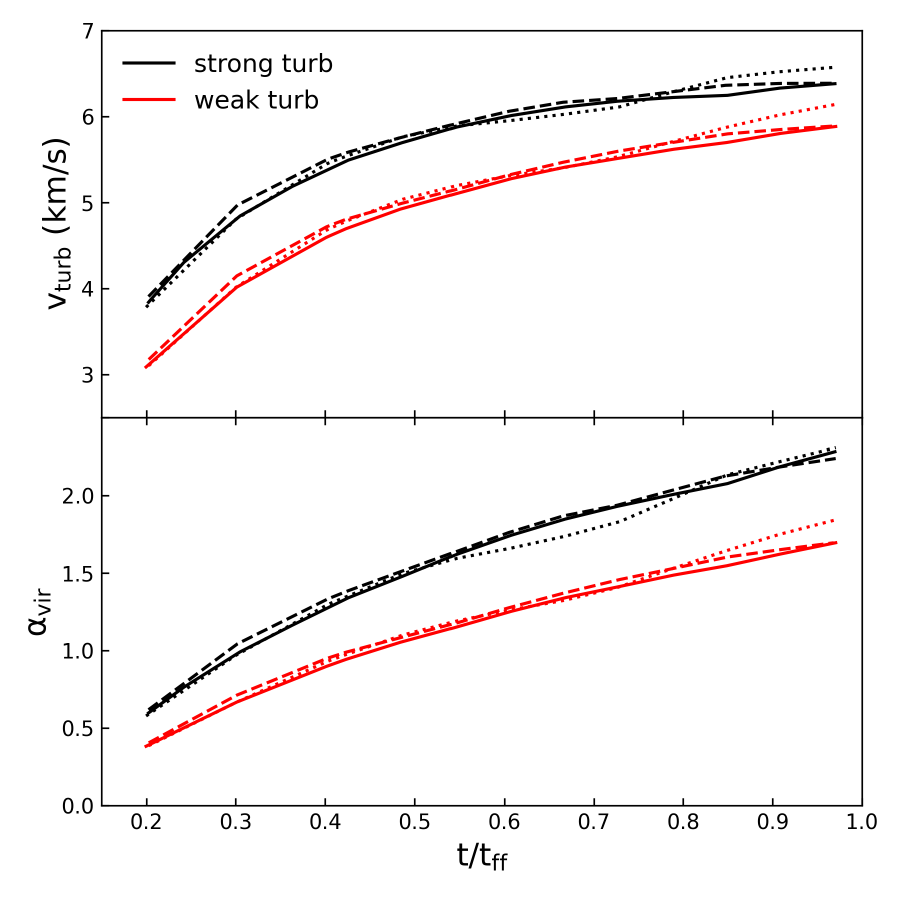}
    \caption{Time evolution of the turbulent velocity (top) and the virial parameter (bottom) measured inside a GMC with a radius of 20 pc. The time is normalized to the free-fall time of the cloud ($\approx 3.3\, \rm Myr$). The black and red lines correspond to the strong and weak turbulence runs with a resolution of 0.0625 pc, respectively. Different line styles represent the runs with different random seeds.}
    \label{fig:GMC_vturb}
\end{figure}

Figure~\ref{fig:SIS_scheme} also shows that the accretion rate based on the Bondi-Hoyle scheme is more accurate than that of the flux accretion method. This is partly due to the fact that we use the total mass of the sink particle and the enclosed gas within the accretion zone, rather than the sink mass alone, to determine the Bondi-Hoyle radius. \citet{Bleuler2014} further showed that even in the regime where $\rsonic \approx 0.1\dxmin$, the spherical Bondi accretion is well reproduced by the Bondi-Hoyle accretion scheme. 
On the other hand, by running AU-scale collapsing sphere simulations with a more realistic outcome, where sink particles form inside an accretion disk, these authors also demonstrated that the density and velocity profiles at the boundary of the accretion zone change violently when the Bondi-Hoyle accretion scheme is used, while the flux accretion method yields smoother profiles. Therefore, following \citet{Bleuler2014}, we adopt the flux accretion scheme when $\rsonic > \racc$, while Bondi-Hoyle accretion is used for the opposite condition.

\begin{figure*}
\centering
    \includegraphics[width=\textwidth]{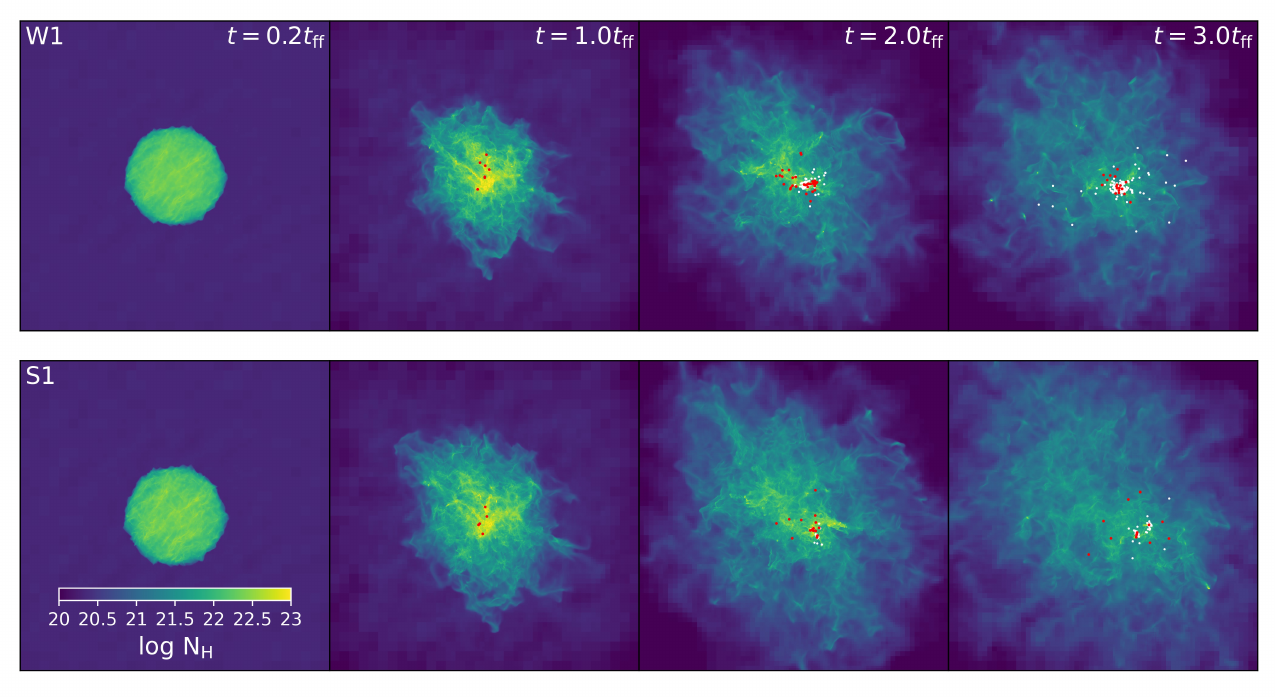}
    \caption{Evolution of GMCs with weak (W1, top) and strong (S1, bottom) turbulence. Different columns show the hydrogen column density distribution at four different epochs. The red and white dots represent sink and star particles, respectively. Each panel measures 128 pc on a side and the maximum resolution of the simulations is 0.0625 pc. No stellar feedback is included.}
    \label{fig:GMC_map}

    \includegraphics[width=\textwidth]{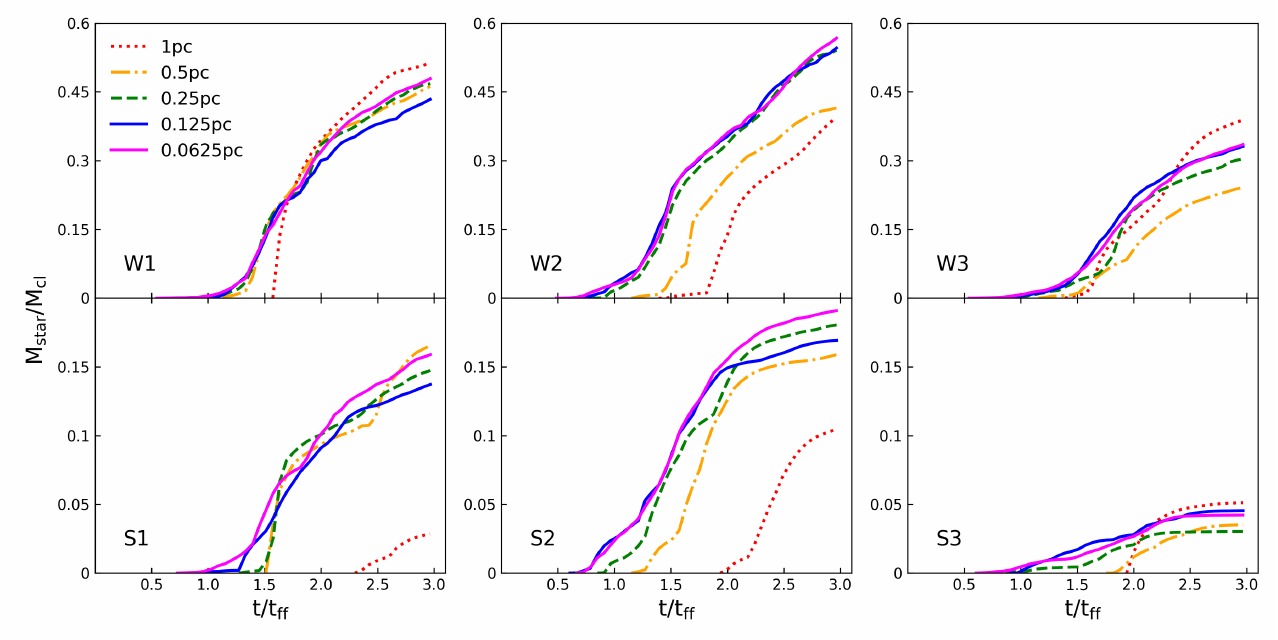}
    \caption{Total SFE ($\mstar/M_{\rm cl}$) in six different GMCs. The clouds with weak and strong turbulence are plotted in the top and bottom panels, respectively. Different columns correspond to simulations with different random seeds for turbulence driving. Simulations with different resolutions are shown in different colors and line styles, as indicated in the legend. The star formation efficiency converges from runs with a resolution of $\sim 0.25\, {\rm pc}$.}
    \label{fig:GMC_SFE}
\end{figure*}

\subsection{Turbulent GMCs}\label{sec:A2}
In this section, we investigate what resolution is required for the sink-based model to reasonably predict stellar mass growth in a turbulent medium, and estimate the model uncertainty. To do this, we use GMC simulations with varying levels of turbulence and compare the formation and growth of sink particles at different resolutions.

\subsubsection{Simulation setup}
The simulated cloud has an initial uniform density of $\nH \approx 186\, \cmq$, with a total mass of $2\times10^5\, \msun$ and a radius of 20 pc, respectively. These values are chosen to mimic the observed properties of typical molecular clouds in the Milky Way \citep{Roman-Duval2010}. The cloud is located at the center of a $(128\, \rm pc)^3$ box with outflowing boundary conditions. The cloud has a uniform temperature of 30 K and a metallicity of $0.1\, Z_\odot$, while the ambient gas temperature outside the cloud is set to $10^4\, \rm K$ to maintain pressure equilibrium with the cloud. The free-fall time and thermal Jeans length of the cloud are $\tff \approx 3.3\, \rm Myr$ and $\lambda_{\rm J} \approx 4.7\, \rm pc$, respectively. The simulation box is divided into $32^3$ root cells, which are further refined if the local thermal Jeans length is not resolved by 32 cells. The maximum resolution is 1 pc for the lowest resolution run and 0.0625 pc for the highest resolution case. Turbulent velocity fields are driven without gravity for the first 0.2 \tff\ so that the Kolmogorov spectrum develops throughout the box. This is done by generating random acceleration field at each time step via Ornstein-Uhlenbeck process with the same turbulence mixing ratio \citep[e.g.,][]{Federrath2010turb,Han2022,Brucy2024} as in Eq.~\eqref{eq:eff}. Self-gravity is then activated while the turbulence is continuously driven.
No stellar feedback is included. We ran a total of six simulations with three different random seeds for the turbulence driving and two different turbulence strengths. These six clouds are labeled S1, S2, and S3 (strong turbulence), and W1, W2, and W3 (weak turbulence), where the same number indicates the same random seed. The density threshold for sink formation is $\rho_{\rm th}=\rho_{\rm LP} \approx8.15\times10^{-21}\left(\dxmin/{1\,\rm pc}\right)^{-2}\, {\rm g \,cm^{-3}}$, or $\approx3730\left(\dxmin/{1\,\rm pc}\right)^{-2}\, \cmq$ in hydrogen number density. Based on the results of Sect.~\ref{sec:A1}, the accretion radius is set to $\racc=2\dxmin$. The mass of each star particle is $100\, \msun$.

The early evolution of the turbulent velocity and the virial parameter of these six clouds is shown in Fig.~\ref{fig:GMC_vturb}. Here, the turbulent velocity $v_{\rm turb}$ is the second moment of the velocity PDF within the sphere, weighted by the cell mass \citep[see e.g., Eqs. (2) and (3) of][]{Ossenkopf2002}. The turbulent Mach number $\mathcal{M}$ ranges from 7--10. Although the virial parameter exceeds 1 and continues to increase even after gravity is turned on, stars form at the local density maxima where the gas structures are locally gravitationally bound. 
This is illustrated in more detail in Fig.~\ref{fig:GMC_map} where the formation of sink and star particles is plotted against the hydrogen column density of two clouds, S1 and W1. Before gravity is turned on ($t=0.2\tff$), the two clouds do not fragment and retain their initial spherical shape with some density contrast due to turbulence. Over time, clumpy structures develop locally, forming sinks and stars, while unbound gas is distributed throughout the box following the net turbulent velocity field. Due to the same random seed for the turbulence driving, the overall structure of the two clouds is similar, but the W1 cloud is more compact, gravitationally bound and thus forms more stars. By $t=3\, \tff$, about three times more stars are created in W1 than in S1.

\subsubsection{Testing the resolution convergence}

\begin{figure}
    \includegraphics[width=\linewidth]{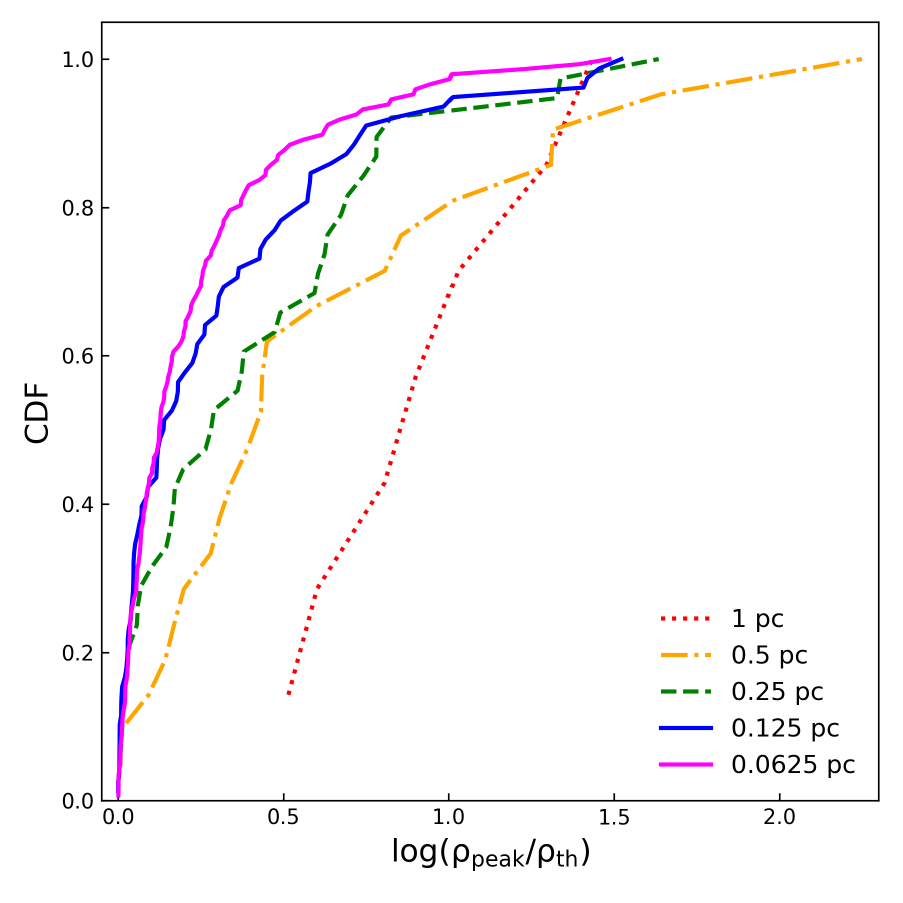}
    \caption{Cumulative distribution functions (CDFs) of the peak density of gas clumps at the time of sink formation ($\rho_{\rm peak}$), normalized to the threshold density for sink creation ($\rho_{\rm th}$). The CDFs are computed by counting all sink particles from strongly turbulent GMCs (S1--S3), where the convergence in the SFEs is worse than in the weakly turbulent GMCs. Different colors and line styles correspond to CDFs from runs with different spatial resolutions, as indicated in the legend. Note that most of the sinks are created near $\rho_{\rm th}$ when the AMR resolution is high.}
    \label{fig:GMC_density_peak}
\end{figure}

\begin{figure}
    \includegraphics[width=\linewidth]{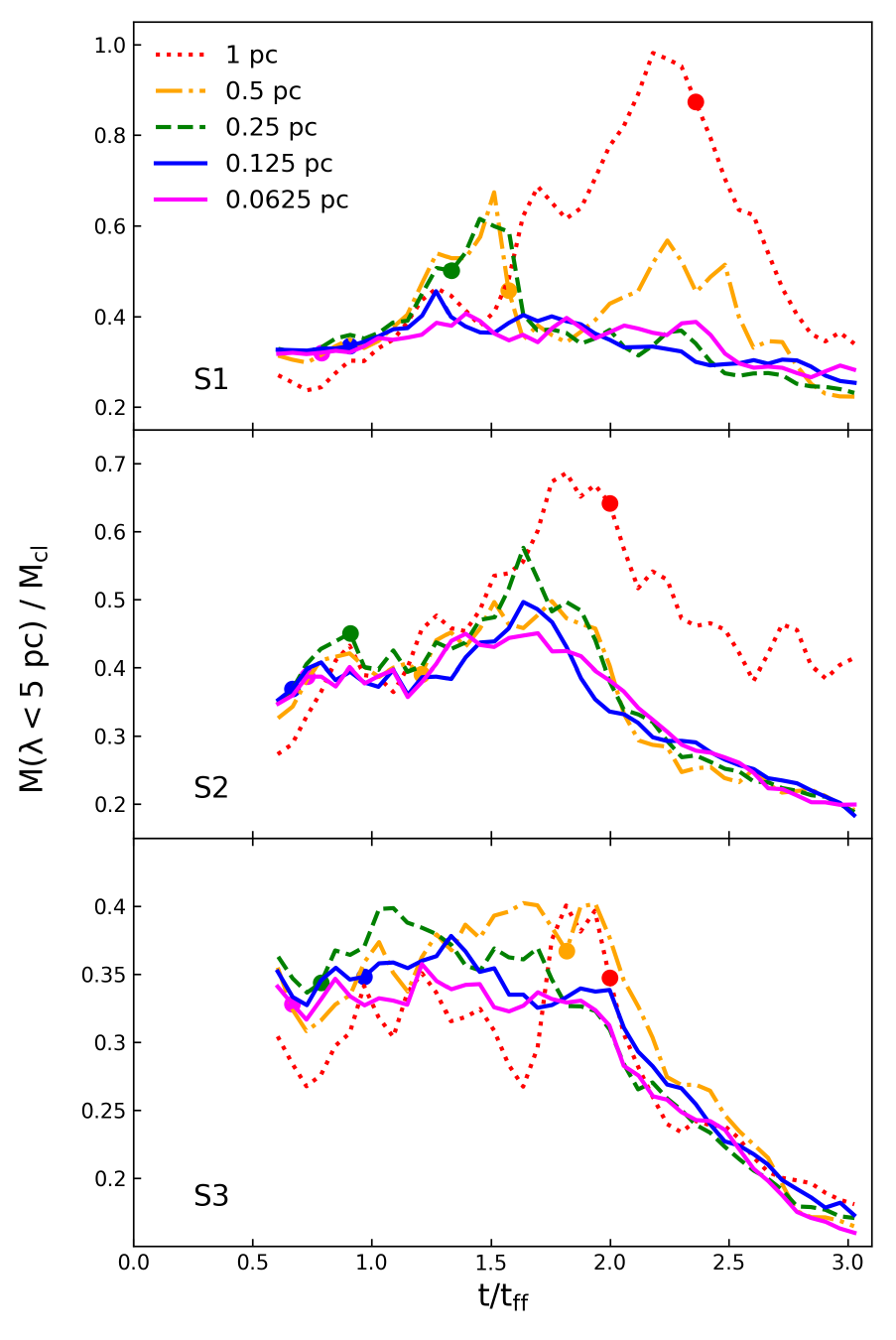}
    \caption{Mass fraction of the gas column density distribution reconstructed with a high pass filter ($\lambda<5\, \rm pc$) to the initial cloud mass ($M_{\rm cl}$). This is done by performing an inverse Fourier transform on the Fourier transformed gas column density map of S1 (top), S2 (middle), and S3 (bottom) after removing signals from scales larger than 5 pc. The filled circle indicates the time at which the first sink particle is produced. The pronounced peaks in the lowest resolution runs of S1 and S2 indicate that the gas is slowly collapsing.}
    \label{fig:GMC_ps5}
\end{figure}

Figure~\ref{fig:GMC_SFE} shows the total stellar mass divided by the initial cloud mass (i.e., the total SFE) in the six different GMCs for five different AMR resolutions. Note that the simulations cover a wide range of the SFE, from $\approx 4\%$ (S3) to $\approx 60\%$ (W2), depending on turbulence strength and random seed. We find that the total SFEs have converged reasonably well for $0.0625 \le \dxmin \le 0.25\,{\rm pc}$ resolutions. In the weak turbulence cases (W1--W3), the self-gravity of the local clumps is stronger than the turbulence, and almost all of the infalling gas in the vicinity of each sink is eventually accreted onto the particle. As a result, a small difference in the turbulent structures and accretion rates in the runs with different resolutions does not significantly affect the overall SFE. Similarly, even when the cloud is broken up into smaller structures by strong turbulence (S1--S3), the final stellar mass at $3\,\tff$ remains relatively similar within $\sim 30\%$ in the $\dxmin \le 0.25\,{\rm pc}$ runs, although it becomes somewhat more sensitive to the resolution. In some cases (W1 and W3) convergence appears to be achieved even in low resolution runs with $\dxmin \sim 1\,{\rm pc}$, but the first star particles are formed later than in the corresponding higher resolution runs by more than $\sim 1\, \tff$. This could have led to dramatically different SFHs if stellar feedback had been included. 

To understand how convergence is achieved, we first examine the formation of sink particles. In our simulations, clumps are defined at local density maxima with densities greater than $0.1\rho_{\rm th}$, and sinks form within these clumps when their peak density increases by an order of magnitude, while maintaining a continuous net inflow and the entire clump remains virialized. To see how sink formation is affected by resolution, we show in Fig.~\ref{fig:GMC_density_peak} the cumulative distribution function of peak densities for clumps from S1--S3, at the epoch of sink formation. There is a clear trend for sink particles to form at densities closer to $\rho_{\rm th}$ in the higher resolution runs. For example, 50\% of the total sink particles form at densities lower than $\approx 1.33\rho_{\rm th}$ in the $0.0625\, \rm pc$ runs, while the clump peak density must increase to $\sim 10\,\rho_{\rm th}$ to form half of the sink particles in the 1 pc runs. This means that resolved clumps are already gravitationally bound or close at $\rho_{\rm th}$, and thus there is only a small difference in the formation time of the first sink. In contrast, the gas clumps in the 1 pc runs are still super-virial even at densities several times higher than $\rho_{\rm th}$, delaying the formation of the sink. At the intermediate resolution (0.5 pc), only in S1 are the formation time of the first sink and the SFHs found to be similar to those in the corresponding higher resolution runs. In S2 and S3 (still at 0.5 pc resolution), a significant fraction ($\sim 20\%$) of the total number of sink particles form at densities higher than $\sim 10\rho_{\rm th}$. These simple experiments suggest that a resolution of 0.25 pc is desirable to resolve the virialized, small scale structures in turbulent GMCs. A lower resolution (0.5 pc) can provide a reasonable estimate of the stellar mass in some cases, but the gravitational binding of gas clumps will in general be more sensitive to the specific properties of turbulence.

As sink particles form and accrete gas in local dense clumps, we also investigate how small scale structures are affected by spatial resolution using Fourier analysis. We first generate a two-dimensional gas column density map for each simulation output by projecting along the $z$ axis, as illustrated in Fig.~\ref{fig:GMC_map}. We then perform a Fourier transform of these gas maps to obtain the power spectrum as a function of wave number.
To focus on the small scale structures relevant to star formation, we remove the signal from scales larger than $\lambda=5\,{\rm pc}$, which is the typical clump size in the 1 pc resolution runs, and reconstruct the map by performing an inverse Fourier transform of the small scale signals only. The ratio of the total mass of the regenerated gas map to the initial cloud mass is shown in Fig.~\ref{fig:GMC_ps5}. Roughly speaking, 40\% of the gas mass in these clouds is concentrated at $\lambda<5\,{\rm pc}$ until the dense gas is dispersed by strong turbulence at $t/\tff \gtrsim 2$. Note that these fractions are at least twice as large as the total star formation efficiencies in S1--S3, and thus include some non star-forming gas. Figure~\ref{fig:GMC_ps5} shows that gas mass fractions in the three high resolution runs are in good agreement. By contrast, in all 1 pc resolution runs the gas fraction is initially smaller and then becomes too large at later times ($t/\tff\sim 2$). This suggests that the clumps are slowly collapsing at lower resolutions while in higher resolution runs where the mass fraction does not increase significantly, sink particles accrete gas more rapidly, removing mass from the dense regions. Taken together with Fig.~\ref{fig:GMC_density_peak}, we conclude that achieving a resolution of at least 0.25 pc is necessary to resolve the small scale structures feeding sink particles.

It is interesting to compare this scale length with the sonic length of the clouds. The theory of supersonic turbulence suggests that the sonic length ($l_{\rm s}$) can be approximated as $l_{\rm s} = l_0\, (\sigma_{l_0}/c_{\rm s})^{-2}$, where $\sigma_{l_0}$ is the velocity dispersion measured on the scale of the cloud diameter ($l_0$) \citep{McKee2007,Federrath2021}. On scales larger than $l_{\rm s}$, supersonic compressive turbulence leads to rapidly varying density fluctuations in space, which in principle must be resolved in simulations to accurately model star formation. To this end, we compute the sonic length in our runs, using the initial diameter of the cloud ($40\, \rm pc$), the mean gas temperature, and the 3D velocity dispersion within the cloud. We find that $l_{\rm s}$ is about 0.2--0.3 pc, indicating that the star-forming structures begin to be resolved in the run with $\dxmin=0.25\,{\rm pc}$, consistent with our interpretation from Fig.~\ref{fig:GMC_density_peak}. Note that $l_{\rm s}$ in our runs is larger than the typical sonic length of GMCs in the Milky Way ($l_{\rm s}\sim0.1\,{\rm pc}$) \citep{Brunt2010,Yun2021}, due to our clouds having a larger temperature ($\sim$ 100 K instead of $\sim$ 10 K). We attribute this difference to both an underestimate of the cooling channels in our reduced chemical networks and the low overall metallicity we assume for our clouds. Follow-up simulations, which we plan to perform in the near future, will include more detailed chemistry \citep[e.g.,][]{Katz2022Prism,Kim2023} to better capture the turbulent structure of star-forming clouds. 

Finally, we compare our results with previous findings. \citet{Truelove1997} showed that the thermal Jeans length should be resolved by at least 4 cells to avoid spurious fragmentation in a collapsing cloud. \citet{Federrath2011} further argued, by simulating a magnetized gas cloud at different resolutions, that the thermal Jeans length should be resolved by 32 cells to obtain convergence of turbulent energies. Since the thermal Jeans length of the cloud in our simulations is $\lambda_{\rm J} \approx 4.7\, \rm pc$, these resolution requirements correspond to $\dxmin \lesssim 1.18\, \rm pc$ and $\dxmin \lesssim 0.15\, \rm pc$, respectively. Our results suggest that an intermediate value of 20 cells per thermal Jeans length is enough to model star formation in turbulent clouds in hydrodynamics simulations. This is not surprising, given that our clouds are more complex than those studied in \citet{Truelove1997} but we ignore magnetic fields. In their MHD simulations of a mini dark matter halo of $\sim 10^6\,\msun$, \citet{Turk2012} claimed that as many as 64 cells per thermal Jeans length are needed to accurately capture the small scale dynamo and hydrodynamic properties of Pop III star-forming clouds. We further discuss resolution effects on star formation histories in a cosmological environment in Sect.~\ref{sec:30}.

To summarize, with the accretion radius fixed at $\racc=2\dxmin$, six different GMCs simulated at five different resolutions from 1 to 0.0625 pc (Fig.~\ref{fig:GMC_map}) show that the stellar mass formed during $t=2\tff$ begins to converge at a resolution of $\sim 0.25\, {\rm pc}$ (Fig.~\ref{fig:GMC_SFE}). Gas clumps in turbulent GMCs (S1) simulated at low resolutions (0.5--1 pc) contract more slowly than their higher resolution run counterparts (Figs.~\ref{fig:GMC_density_peak}--\ref{fig:GMC_ps5}), leading to delayed formation of sink particles and lower overall stellar masses.

\label{LastPage}
\end{document}